\documentclass[aos, preprint]{imsart}

\RequirePackage[OT1]{fontenc}
\RequirePackage{amsthm,amsmath}
\allowdisplaybreaks[4]
\usepackage{array}
\usepackage{mathrsfs}
\usepackage{arydshln}
\usepackage{amssymb,bbm}
\usepackage{amsfonts}
\usepackage{amsmath,amssymb}
\usepackage{graphicx}
\usepackage{amscd}
\usepackage{epstopdf}
\usepackage{enumerate}
\usepackage{color}
\usepackage{mathrsfs}
\usepackage{latexsym,bm}
\usepackage{array}
\usepackage{threeparttable}

\usepackage{threeparttable}
\usepackage{rotating}
\usepackage{booktabs}

\usepackage{rotating}
\usepackage{multirow}
\usepackage{amsmath}
\usepackage{subfigure}

\usepackage{makecell}
\usepackage{subfigure}
\usepackage{diagbox}
\usepackage{enumitem}
\usepackage{bm}

\usepackage[plain,noend]{algorithm2e}

\DeclareFontFamily{U}{mathx}{\hyphenchar\font45}
\DeclareFontShape{U}{mathx}{m}{n}{
	<5> <6> <7> <8> <9> <10>
	<10.95> <12> <14.4> <17.28> <20.74> <24.88>
	mathx10
}{}
\DeclareSymbolFont{mathx}{U}{mathx}{m}{n}
\DeclareMathAccent{\widecheck}{\mathalpha}{mathx}{"71}

\DeclareMathAccent{\widecheck}{\mathalpha}{mathx}{"71}

\startlocaldefs \numberwithin{equation}{section} \theoremstyle{it}
\newtheorem{thm}{Theorem}[section]

\newtheorem{de}{Definition}[section]
\newtheorem{ass}{Assumption}

\newtheorem{pro}{Proposition}[section]

\endlocaldefs

\newcommand{\f}{\frac}

\begin{document}

	\begin{frontmatter}
		\title{Multi-frequency-band tests for white noise under heteroskedasticity}
		
		\begin{aug}
			\author{\fnms{Mengya} \snm{Liu}\thanksref{m1}\ead[label=e1]{???}},
			\author{\fnms{Fukang} \snm{Zhu}\thanksref{m1}
            \ead[label=e3]{???}}
			\and
			\author{\fnms{Ke} \snm{Zhu}\thanksref{m3}
				\ead[label=e4]{???}
				\ead[label=u1,url]{http://www.foo.com}}

			
			\affiliation{Jilin University\thanksmark{m1} and University of Hong Kong\thanksmark{m3}}
			%
			%
			%
			%
		\end{aug}
	
\begin{abstract}
This paper proposes a new family of multi-frequency-band (MFB) tests for the white noise hypothesis by using the maximum overlap discrete wavelet packet transform (MODWPT).
The MODWPT allows the variance of a process to be decomposed into the variance of its components on different equal-length frequency sub-bands, and the MFB tests then measure the
distance between the MODWPT-based variance ratio and its theoretical null value jointly over several frequency sub-bands. The resulting MFB tests have the chi-squared asymptotic null distributions under mild conditions, which allow the data to be heteroskedastic. The MFB tests are shown to have the desirable size and power performance by simulation studies, and their usefulness is further illustrated by two applications.
\end{abstract}


\begin{keyword}
\kwd{Heteroskedasticity; Maximum overlap discrete wavelet packet transform; Testing for white noise; Variance ratio test; Wavelets} 
\end{keyword}
\end{frontmatter}

\newpage
\section{Introduction}

Consider a stochastic sequence $\{y_t\}$ with $E(y_t)=0$ for all $t\in\mathbb{Z}$.
A long standing problem in time series analysis is to detect the null hypothesis that $\{y_t\}$ is  white noise, i.e.,
\begin{equation}\label{null_hypo}
H_0:\, \{y_t\} \mbox{ is an uncorrelated process}.
\end{equation}
In the time domain, Box and Pierce (1970) and later Ljung and Box (1978) proposed portmanteau tests to detect $H_0$ by checking whether
$E(y_t y_{t-k})=0$ at some finite lags $k=1,...,K$. Their portmanteau tests require $\{y_t\}$ to be independent and
identically distributed (i.i.d.), while the i.i.d. condition is restrictive in many economic and financial applications. To relax this condition, Lobato, Nankervis and Savin (2001)
constructed a modified portmanteau test, which is valid when $\{y_t\}$ is a martingale difference sequence (MDS). This method was
further studied by Escanciano and Lobato (2009) with a data-driven method to select an optimal lag.
For the non-MDS $\{y_t\}$, some robust versions of portmanteau test were proposed in
Romano and Thombs (1996) and Horowitz, Lobato, Nankervis and Savin (2006) by implementing the block bootstrap methods,
Lobato (2001) by using the self-normalization technique, and Lobato, Nankervis and Savin (2002) and Zhu (2016)
by estimating the asymptotic variance matrix of the first $K$ sample autocorrelations of $\{y_t\}$.
However, all of the aforementioned tests require $\{y_t\}$ to be stationary, and they are thus
 not applicable for heteroskedastic $\{y_t\}$ (i.e., $Ey_t^2\not\equiv$ a constant for all $t$).

\begin{figure}[!h]
	\includegraphics[width=36pc]{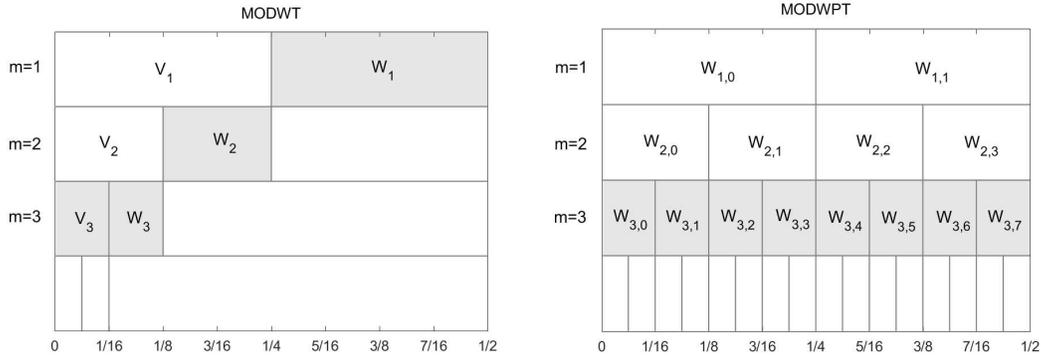}
	\caption{The decomposition ways of MODWT (left) and MODWPT (right). For the MODWT, only $V_m$ at scale $m$ is decomposed into
$V_{m+1}$ and $W_{m+1}$ at scale $m+1$. For the MODWPT, all $\{W_{m,n}\}_{n=0}^{2^{m}-1}$ at scale $m$ are decomposed into
$\{W_{m+1,n}\}_{n=0}^{2^{m+1}-1}$ at scale $m+1$.}\label{M}
\end{figure}

In the frequency domain, Gen\c{c}ay and Signori (2015) recently introduced a family of
multi-scale tests for $H_0$, and their tests work for the heteroskedastic $\{y_t\}$.
To illustrate the idea of multi-scale tests, we simply assume that $\{y_t\}$ is a covariance stationary process.
The multi-scale tests first apply the maximum overlap discrete wavelet transform (MODWT) to $\{y_t\}$, and then
obtain its high frequency component $W_m\equiv\{W_{m,t}\}$ and low frequency component $V_m\equiv\{V_{m,t}\}$ at each scale $m$, where
$W_m$ and $V_m$ are related to the frequency sub-bands
$[\frac{1}{2^{m+1}}, \frac{1}{2^m}]$ and $[0,\frac{1}{2^{m+1}}]$, respectively, and they are decomposed recursively from $V_{m-1}$; see the left panel in Figure\,\ref{M}
for the decomposition way of MODWT.
Next, Gen\c{c}ay and Signori (2015) showed that if $\{y_t\}$ is  white noise,
\begin{equation}\label{modwt_wvr}
\frac{\mbox{var}(W_{m,t})}{\mbox{var}(y_t)}=\frac{1}{2^m} \,\,\,\mbox{ for }m=1,2,...,
\end{equation}
where $\mbox{var}(W_{m,t})$ is the MODWT-based wavelet variance, and so $\mbox{var}(W_{m,t})/\mbox{var}(y_t)$
is the MODWT-based wavelet variance ratio (WVR).
Motivated by (\ref{modwt_wvr}), the multi-scale tests detect $H_0$
by measuring the distance (under certain norm) between the sample version of MODWT-based WVR and $\frac{1}{2^m}$
at each scale $m$ (or jointly over the first $m$ scales).
With the aid of wavelet method, the multi-scale tests
are particularly suitable in situations where the data $\{y_t\}$ have
jumps, kinks, seasonality and non-stationary features. This advantage does not hold for the Fourier-based
frequency-domain tests in Hong (1996), Paparoditis (2000), Fan and Zhang (2004), Escanciano and Velasco (2006),
 and Shao (2011a).
Besides the multi-scale tests, some other wavelet-based frequency-domain tests were
constructed based on the wavelet spectral density estimator. In this context,
Lee and Hong (2001) applied the idea of Hong (1996) to construct an asymptotically pivotal test, but their test requires $\{y_t\}$ to be
stationary and homoskedastic, and its result is usually sensitive to the choice of the finest scale
especially when the sample size is small; Duchesne, Li and Vandermeerschen (2010) and Li, Yao and Duchesne (2014) further
developed some wavelet-based tests by using the idea of Fan (1996), however, their methods are only applicable
for the stationary i.i.d. data,  with some bootstrap methods to obtain the critical values.

\begin{figure}[!h]
	\includegraphics[width=34pc]{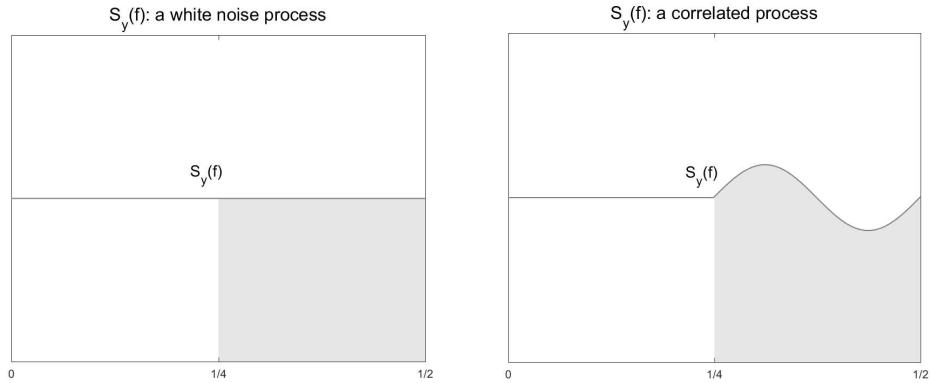}
	\caption{The plot of $S_y(f)$ for a white noise process (left) and a correlated process (right).
The contribution of frequency band [1/4,1/2] to the total variance of each process is in gray.}\label{var}
\end{figure}

Although the multi-scale tests have the aforementioned advantage over the existing ones, they have a drawback due to
the decomposition way of MODWT. To see it clearly, we note that
for any covariance stationary process $\{y_t\}$ and $m=1,2,...$,
\begin{equation}\label{relation_wvr}
\frac{\mbox{var}(W_{m,t})}{\mbox{var}(y_t)}\approx\frac{\int_{1/2^{m+1}}^{1/2^m}S_{y}(f)df}{\int_{0}^{1/2}S_{y}(f)df}
\end{equation}
(see Gen\c{c}ay and Signori (2015)), where
$S_y(f)$ is the spectral density function of $\{y_t\}$, and it is flat under $H_0$. The result (\ref{relation_wvr}) implies that the MODWT-based WVR at scale $m$ essentially measures the ratio of the total variance contributed by the frequency sub-band $[\frac{1}{2^{m+1}}, \frac{1}{2^m}]$.
So, the multi-scale tests lack the power if $S_y(f)$ is not flat but satisfies the relationship:
\begin{equation*}
\frac{\int_{1/2^{m+1}}^{1/2^m}S_{y}(f)df}{\int_{0}^{1/2}S_{y}(f)df}\approx\frac{1}{2^m}\,\,\,\mbox{ for }m=1,2,....
\end{equation*}
As a simple illustrating example, Figure\,\ref{var} plots $S_y(f)$ for a white noise process  and
a correlated process. By construction,
the contribution of frequency sub-band $[\frac{1}{2^{m+1}}, \frac{1}{2^m}]$ to the total variance of each process is the same, and
 the multi-scale tests  are thus unable to distinguish these two processes.
To detect this correlated process, an intuitive way is to further decompose the high-frequency component $W_{m}$, so that
more signals to reject $H_0$ can be found within the frequency sub-band $[\frac{1}{2^{m+1}}, \frac{1}{2^m}]$. However, the MODWT fails to do this, since it does not re-decompose $W_{m}$ any more.

This paper is motivated to propose a new family of frequency-domain-based tests for $H_0$
by using the maximum overlap discrete wavelet packet transform (MODWPT).
The MODWPT decomposes the process $\{y_t\}$ into
$2^{m}$ different components $\{W_{m,n}; n=0,...,2^{m}-1\}$ at each scale $m$, where
$W_{m,n}\equiv \{W_{m,n,t}\}$ is related to the frequency sub-band $[\frac{n}{2^{m+1}}, \frac{n+1}{2^{m+1}}]$,
and it is decomposed recursively from $2^{m-1}$ components $\{W_{m-1,n}\}$ at the previous scale;
see the right panel in Figure\,\ref{M} for the decomposition way of MODWPT. Unlike the MODWT,
the MODWPT re-composes each  $W_{m,n}$ so that the entire frequency band $[0,\frac{1}{2}]$ is refined,
and it thus provides us with an effective way to largely overcome the inconsistency problem
in multi-scale tests. With $\{W_{m,n}; n=0,...,2^{m}-1\}$, our testing principle uses the fact that if $\{y_t\}$ is stationary white noise,
\begin{equation}\label{modwpt_wvr}
\frac{\mbox{var}(W_{m,n,t})}{\mbox{var}(y_t)}=\frac{1}{2^m} \,\,\mbox{ for }n=0,...,2^{m}-1,
\end{equation}
where $\mbox{var}(W_{m,n,t})$ is the MODWPT-based wavelet variance, and $\mbox{var}(W_{m,n,t})/\mbox{var}(y_t)$
is the MODWPT-based WVR.
Hence, at each scale $m$, we can look for the rejection evidence by
measuring the distance between the sample version of MODWPT-based WVR and $\frac{1}{2^m}$
jointly over $n=1,...,2^{m}-1$. Note that we do not consider the testing signal in $W_{m,0}$ (which is identical to $V_{m}$) as done in
Gen\c{c}ay and Signori (2015).
Our resulting tests are called the multi-frequency-band (MFB) tests, since they are constructed
by collecting signals from all frequency sub-bands (except the first one) at each scale $m$.
The MFB tests are shown to have simple chi-squared limiting null distributions,
under conditions that allow for higher order dependence, heteroskedasticity, and trending moments.
Hence, they are easy-to-implement with great generality.
Simulation studies show that the MFB tests can have desirable empirical size and power even when the sample size is small, and they
can perform better than the multi-scale tests and other competitors especially when the serial dependence of the examined data exists at large lags.
Also, the simulation studies indicate that the multi-scale tests could serve as diagnostic tools for many
non-stationary models, including, for example, the time-varying GARCH model in Subba Rao (2006), the non-stationary GARCH model in Francq and Zako\"{i}an (2012),
and the ZD-GARCH model in Li, Zhang, Zhu and Ling (2018), whose model diagnostic checking methods are absent in the literature.

Finally, two applications are given to demonstrate the usefulness of the MFB tests.
In the first application, our MFB tests show that although the entire S\&P500 return series in 2006--2015 is not white noise,
its sub-series in 2009--2015 is white noise. These results are informative for empirical researchers, since they indicate that
the S\&P500 stock market possibly is not predictable in 2009--2015 but predictable in 2006--2008. Since the S\&P500 stock market is
relatively more volatile in 2006--2008 than 2009--2015, our findings may suggest that the S\&P500 stock market is more likely to be inefficient when it is more volatile.
In the second application, we apply our MFB tests to four non-stationary stock return series in Francq and Zako\"{i}an (2012), and find that three of them are not white noises.
Hence, it implies that these three non-white-noise series have some dynamical structures in their conditional mean, and they should not be directly fitted by the first-order
non-stationary GARCH model as done in Francq and Zako\"{i}an (2012).

The remainder of this paper is organized as follows. Section 2 introduces the MODWPT-based WVR and gives the asymptotics of its
estimator. Section 3 proposes our MFB tests and studies their asymptotics.
Simulations are provided in Section 4 and applications are offered in Section 5.
Technical proofs are deferred to the Appendix.


\section{Wavelet variance ratio and its estimator}
The wavelet variance ratio (WVR) plays an important role in our testing principle.
Below, we introduce the WVR based on
the maximum overlap discrete wavelet packet transform (MODWPT) and its estimator.
For more discussions on MODWPT, we refer to Percival and Walden (2000).

\subsection{MODWPT-based WVR}
To elaborate the definition of the MODWPT-based WVR, we simply assume that
$\{y_t\}_{t=1}^{T}$ is a stationary process with mean zero.
The MODWPT-based WVR is defined in terms of the MODWPT component of $\{y_t\}_{t=1}^{T}$.
To compute the MODWPT component, we need a wavelet filter $\{h_l\}_{l=0}^{L-1}$ and its associated scaling filter $\{g_l\}_{l=0}^{L-1}$, where
$\{h_l\}_{l=0}^{L-1}$ satisfies that $h_l=0$ for $l<0$ or $l\geq L$, and
\begin{align*}
\sum_{l=0}^{L-1}h_l=0,\
\sum_{l=0}^{L-1}h^2_l=1,\
\sum_{l=-\infty}^{\infty}h_lh_{l+2n}=0,
\end{align*}
and $\{g_l\}_{l=0}^{L-1}$ satisfies that $g_l=(-1)^{l+1}h_{L-1-l}$ and
\begin{align*}
\sum_{l=0}^{L-1}g_l=1,\ \sum_{l=0}^{L-1}g^2_l=1,\ \sum_{l=-\infty}^{\infty}g_lg_{l+2n}=0,\ \sum_{l=-\infty}^{\infty}g_lh_{l+2n}=0,
\end{align*}
for all nonzero integers $n$. Some well-known choices of $h_l$ and $g_l$ are given as follows:
\begin{itemize}
	\item Haar wavelet: $\{h_l\}_{l=0}^{1}=(1/2, -1/2)$ and  $\{g_l\}_{l=0}^{1}=(1/2, 1/2)$.

	\item Daubechies wavelets ($D(L)$): $D(2)$ is
	just the Haar wavelet. The wavelet and scaling filters for $D(4)$ are defined as
	$$\{h_l\}_{l=0}^{3}=\left( \frac{1-\sqrt{3}}{8},\frac{-3+\sqrt{3}}{8}, \frac{3+\sqrt{3}}{8},\frac{-1-\sqrt{3}}{8}\right) $$ and
	$$\{g_l\}_{l=0}^{3}=\left( \frac{1+\sqrt{3}}{8},\frac{3+\sqrt{3}}{8},\frac{3-\sqrt{3}}{8},\frac{1-\sqrt{3}}{8}\right),$$respectively.
	 The wavelet and scaling filters for $D(L)$ with $L>4$ can be found in Daubechies (1992).
\end{itemize}

Let $L_m=(2^m-1)(L-1)+1$ for some integer $m\geq 1$. Based on $\{h_l\}_{l=0}^{L-1}$ and $\{g_l\}_{l=0}^{L-1}$, we then compute $\{\widetilde{v}_{m,n,l}\}_{l=0}^{L_m-1}$ by
$$\widetilde{v}_{m,n,l}=\frac{1}{2^{m/2}}v_{m,n,l}$$
for $n=0, 1, ..., 2^{m}-1$. Here, $v_{m,n,l}$ is defined recursively by
\begin{equation*}
v_{m,n,l}=\sum^{L-1}_{k=0}u_{n,k}v_{m-1,\left[ \frac{n}{2}\right],l-2^{m-1}k}
\end{equation*}
with $v_{1,0,l}=g_l$ and $v_{1,1,l}=h_l$, where $\left[\cdot\right]$ is the integer part operator, and
\begin{equation*}
u_{n,l}=\left\{\begin{array}{ll}
g_l,&$if$\ n\bmod 4=0\ $or$\ 3,\\
h_l,&$if$\ n\bmod 4=1\ $or$\ 2.
\end{array}\right.
\end{equation*}
Using $\{\widetilde{v}_{m,n,l}\}_{l=0}^{L_m-1}$, the MODWPT components $W_{m,n}\equiv\{W_{m,n,t}\}_{t=1}^{T}$
at scale $m$ are
computed with the MODWPT coefficients
\begin{equation*}
W_{m,n,t}=\sum^{L_m-1}_{l=0}\widetilde{v}_{m,n,l}y_{t-l \bmod T}.
\end{equation*}
Note that $W_{m,n,t}$ can be fast calculated by using the R package ``wmtsa''.
Generally speaking, the MODWPT at each scale $m$ decomposes the entire frequency band $[0, \frac{1}{2}]$ into $2^m$ equal sub-bands (see the right panel in Figure\,\ref{M}), and the resulting
$W_{m,n}$ contains the characteristics of the original time
series $\{y_t\}_{t=1}^{T}$ in each sub-band $[\frac{n}{2^{m+1}},\frac{n+1}{2^{m+1}}]$.

Similar to Gen\c{c}ay and Signori (2015), we next define the wavelet variance of $\{y_t\}$ in the frequency sub-band $[\frac{n}{2^{m+1}},\frac{n+1}{2^{m+1}}]$ by
\begin{equation}\label{2e2}
{\rm wvar}_{m,n}(y)\equiv{\rm var}(W_{m,n,t}).
\end{equation}
With $\{{\rm wvar}_{m,n}(y)\}$, we can approximately decompose
the variance of $\{y_t\}$ at scale $m$ by
\begin{equation}\label{approxi}
{\rm var}(y)\approx \sum_{n=0}^{2^m-1}{\rm wvar}_{m,n}(y),
\end{equation}
where the result (\ref{approxi}) holds, because
${\rm wvar}_{m,n}(y)\approx {\rm var}_{m,n}(y)\equiv 2\int^{\frac{n+1}{2^{m+1}}}_{\frac{n}{2^{m+1}}}S_y(f)\mathrm{d}f$
by neglecting the leakage of the wavelet filter (see Gen\c{c}ay and Signori (2015)), and
${\rm var}(y)=2\int^{1/2}_{0}S_y(f)\mathrm{d}f=\sum_{n=0}^{2^m-1}{\rm var}_{m,n}(y).$
Here, $S_y(f)$ is the spectral density function of $\{y_t\}$, and ${\rm var}_{m,n}(y)$ can be viewed as
 the general variance of $\{y_t\}$ in the sub-band $[\frac{n}{2^{m+1}},\frac{n+1}{2^{m+1}}]$.

Now, we define the MODWPT-based WVR in the frequency sub-band $[\frac{n}{2^{m+1}},\frac{n+1}{2^{m+1}}]$ by
\begin{equation}\label{WVR}
\xi_{m,n}(y)\equiv\f{{\rm wvar}_{m,n}(y)}{{\rm var}(y)}.
\end{equation}
Clearly, the result (\ref{approxi}) implies that for the general stationary process $\{y_t\}$,
$\sum_{n=0}^{2^m-1}\xi_{m,n}(y)\approx 1.$
Particularly, if $\{y_t\}$ is covariance stationary white noise, Theorem \ref{t1} below shows that the approximation symbol ``$\approx$''
can be replaced by the equality symbol ``$=$''.

\begin{thm}\label{t1}
Suppose $\{y_t\}$ is  covariance stationary white noise. Then,
\begin{equation*}
\xi_{m,n}(y)=\f{1}{2^m}
\end{equation*}
at each scale $m$, where $n=0,...,2^m-1$.
\end{thm}

The preceding theorem demonstrates that if $\{y_t\}$ is covariance stationary white noise,
the MODWPT-based wavelet variance at each sub-band $[\frac{n}{2^{m+1}},\frac{n+1}{2^{m+1}}]$ contributes a ratio of $\frac{1}{2^{m}}$ to the total variance.
In the next section, we will apply this result to form a class of tests for $H_0$.
Specifically, we will measure the distance between $\xi_{m,n}(y)$ and $\frac{1}{2^m}$ under certain norm, and a large value of
 this distance conveys  the evidence of rejection for $H_0$.

\subsection{The estimator of $\xi_{m,n}(y)$}

To facilitate our testing idea, an estimator of $\xi_{m,n}(y)$ is needed. In this paper, we
estimate $\xi_{m,n}(y)$ by $\widehat{\xi}_{m,n,T}$, where
\begin{equation}\label{2e4}
\widehat{\xi}_{m,n,T}=\f{\widehat{{\rm wvar_{m,n}}(y)}}{\widehat{{\rm var}(y)}}\equiv\f{\sum^{T}_{t=1}W^2_{m,n,t}}{\sum^{T}_{t=1}y^2_t}.
\end{equation}

Let $z_{m,n,t}=\sum_{i=0}^{L_m-1}\sum_{j>i}^{L_m}\widetilde{v}_{m,n,i}\widetilde{v}_{m,n,j}y_{t-i}y_{t-j}$ and
\begin{equation}\label{s_mnT}
s_{m,n,T}^2(z)=\frac{1}{T}\sum_{t=1}^T {\rm var}(z_{m,n,t})+\frac{2}{T}\sum_{t=1}^T\sum_{k=1}^{T-1}{\rm cov}(z_{m,n,t},z_{m,n,t-k}),
\end{equation}
where $s_{m,n,T}^2(z)$ is the long run variance of $\frac{1}{\sqrt{T}}\sum_{t=1}^{T}z_{m,n,t}$.
Theorem \ref{t2} below shows that the consistency and asymptotic normality of $\widehat{\xi}_{m,n,T}$ hold even for the
heteroskedastic white noise $\{y_t\}$.

\begin{thm}\label{t2}
Suppose $\{y_t\}$ is heteroskedastic white noise.
For any given $m\geq 1$ and $n=1,...,2^{m}-1$, (i) if Assumption \ref{ass1} in the Appendix holds,
\begin{equation*}
\widehat{\xi}_{m,n,T}\xrightarrow{p}\f{1}{2^m}\mbox{ as }T\to\infty;
	\end{equation*}
(ii) if $\lim_{T\to\infty}\frac{1}{T}\sum_{t=1}^{T}Ey_t^2=\sigma^2<\infty$ and Assumption \ref{ass2} in the Appendix holds,
	\begin{equation} \label{t2mv}
	WV_{m,n}\equiv\sqrt{\frac{T\sigma^4}{4{\rm avar}(z_{m,n})}}\left(\widehat{\xi}_{m,n,T}-\frac{1}{2^m}\right)\xrightarrow{d} N(0,1)\mbox{ as }T\to\infty,
	\end{equation}
where ${\rm avar}(z_{m,n})$ is the probability limit of  $s_{m,n,T}^2(z)$ in (\ref{s_mnT}).
\end{thm}

To implement Theorem \ref{t2}(ii), we need either estimate $\sigma^2$ and ${\rm avar}(z_{m,n})$ consistently or calculate them explicitly.
For the general cases, $\sigma^2$ can be consistently estimated by $\widehat{\sigma}^{2}\equiv \frac{1}{T}\sum_{t=1}^{T}y_t^2$ under some mixingale conditions in Andrews (1988),
and ${\rm avar}(z_{m,n})$ can be consistently estimated by the conventional Newey--West (NW) estimator $\widehat{{\rm avar}}(z_{m,n})$.
For a special case that
\begin{equation}\label{cross_cumulant}
\mbox{all cross-joint cumulants of order four for } \{y_t\} \mbox{ are zeros},
\end{equation}
we can show that $4\sigma^{-4}{\rm avar}(z_{m,n})$ in (\ref{t2mv}) has an explicit formula, which can be directly calculated
from the wavelet filter $\{h_l\}$. Here, the
cross-joint cumulants of order four for $\{y_t\}$ is defined as the coefficients $\kappa^{a,b,c,d}$  in the Taylor's expansion:
$$\log M(\xi)=\sum_{a}\xi_{a}\kappa^{a}+\frac{1}{2!}\sum_{a,b}\xi_{a}\xi_{b}\kappa^{a,b}+\frac{1}{3!}\sum_{a,b,c}\xi_{a}\xi_{b}\xi_{c}\kappa^{a,b,c}
+\frac{1}{4!}\sum_{a,b,c,d}\xi_{a}\xi_{b}\xi_{c}\xi_{d}\kappa^{a,b,c,d}+\cdots,$$
where $M(\xi)=E\exp(\xi'y_{t}^{ijkl})$ with $\xi\in\mathcal{R}^{4\times1}$ and $y_{t}^{ijkl}=(y_{t-i},y_{t-j},y_{t-k},y_{t-l})'\in\mathcal{R}^{4\times1}$ for any
 $i,j,k,l$, and each index in the summation is running from 1 to 4.

\begin{pro}\label{p1}
Suppose $\{y_t\}$ is  heteroskedastic white noise and the condition (\ref{cross_cumulant}) holds. Then,
$WV_{m,n}$ defined in (\ref{t2mv}) can be simplified as
\begin{equation} \label{pro1.1}
WV_{m,n}=\sqrt{\frac{T}{a(\widetilde{v}_{m,n,n})}}\left(\widehat{\xi}_{m,n,T}-\frac{1}{2^m}\right), 
\end{equation}
where
\begin{equation*}
a(\widetilde{v}_{m,n_1,n_2})=\sum_{s\in\textbf{Z}}\,\sum_{i=i_{\min}}^{i_{\max}}\sum_{j\geq i}^{j_{\max}}\widetilde{v}_{m,n_1,i}\widetilde{v}_{m,n_1,j}\widetilde{v}_{m,n_2,i-s}\widetilde{v}_{m,n_2,j-s}
\end{equation*}
with $i_{\min}=\max\{0,s\}$, $i_{\max}=\min\{L_m,L_m+s\}-2$ and $j_{\max}=\min\{L_m,L_m+s\}-1$.
\end{pro}

Note that $WV_{m,n}$ aims to convey the testing signal expressed by the WODWPT-based WVR
within the frequency sub-band $[\frac{n}{2^{m+1}}, \frac{n+1}{2^{m+1}}]$, and
the results of $WV_{m,n}$ in Theorem \ref{t2}(ii) and Proposition \ref{p1}
are key to form our test statistics below.

\section{Multi-frequency-band tests}
In this section, we propose some new test statistics based on the WODWPT-based WVR to detect
the  null hypothesis $H_0$ in (\ref{null_hypo}).
Let ${\bf W}_{m}\equiv(WV_{m,1},\cdots,WV_{m,2^{m}-1})'\in\mathcal{R}^{(2^{m}-1)\times1}$, and
$\Sigma_{m}\in\mathcal{R}^{(2^{m}-1)\times (2^{m}-1)}$ be the asymptotic covariance matrix of ${\bf W}_{m}$ under $H_0$ with its
$(i,j)$th entry
$$\Sigma_{m,i,j}=\frac{{\rm acov}(z_{m,i}z_{m,j})}{\sqrt{{\rm avar}(z_{m,i})}\sqrt{{\rm avar}(z_{m,j})}},$$
where ${\rm acov}(z_{m,i}z_{m,j})$ is the probability limit of the long run covariance of $\frac{1}{\sqrt{T}}\sum_{t=1}^{T}z_{m,i,t}$
and $\frac{1}{\sqrt{T}}\sum_{t=1}^{T}z_{m,j,t}$. Since our testing principle is to measure the distance between $\widehat{\xi}_{m,n,T}$ and $\frac{1}{2^m}$ for $n=1,...,2^m-1$,
a straightforward way is to consider a joint multi-frequency-band test statistic:
\begin{equation}\label{jmf}
MFB_{m}\equiv{\bf W}_{m}^{'}\Sigma_{m}^{-1}{\bf W}_{m}
\end{equation}
at each scale $m$. By construction, we know that under $H_0$,
$$MFB_{m}\xrightarrow{d} \chi^{2}_{2^m-1} \mbox{ as }T\to\infty.$$

Our test $MFB_{m}$ is similar to the multi-scale test $GSM_m$ based on the maximum overlap discrete wavelet transform (MODWT)
in Gen\c{c}ay and Signori (2015),
where
$$GSM_{m}\equiv (GS_1,...,GS_m)\dot{\Sigma}_m^{-1}(GS_1,...,GS_m)',$$
and under $H_0$, $GSM_{m}\xrightarrow{d} \chi^{2}_{m}$ as $T\to\infty$.
Here, $\dot{\Sigma}_{m}\in\mathcal{R}^{m\times m}$ is the asymptotic covariance matrix of $(GS_1,...,GS_m)$ with
$$GS_{m}\equiv \sqrt{\frac{T\sigma^4}{4{\rm avar}(z_{m})}}\left(\widehat{\xi}_{m,T}-\frac{1}{2^m}\right),$$
where $\widehat{\xi}_{m,T}$ is defined as $\widehat{\xi}_{m,n,T}$ in (\ref{2e4}) with $W_{m,n,t}$ replaced by
$W_{m,t}$, ${\rm avar}(z_{m})$ is defined as ${\rm avar}(z_{m,n})$ in Theorem \ref{t2} with $z_{m,n,t}$ replaced by $z_{m,t}^{*}$, and
$$z_{m,t}^{*}=\sum_{i=0}^{L_m-1}\sum_{j>i}^{L_m}h_{m,i}h_{m,j}y_{t-i}y_{t-j}.$$
Like $GSM_{m}$, $MFB_{m}$ can also consistently detect any finite ARMA alternatives and have non-trivial power to detect the local alternative
of the form:
$$H_{1T}: S_{T}(f)=\frac{1}{\sqrt{T}}\Big(S(f)-\frac{1}{2}\Big)+\frac{1}{2},$$
by using the similar arguments as in Gen\c{c}ay and Signori (2015), where $S(f)$ is the non-constant spectrum.
However, the two tests have distinctions due to the different decomposition ways of MODWT and MODWPT as shown in Figure\,\ref{M}.
Specifically, $GSM_m$ looks for the rejection evidence from
the components $\{W_{1},...,W_{m}\}$ at the first $m$ scales, while
$MFB_{m}$ does it from the components $\{W_{m,1},...,W_{m,2^{m}-1}\}$ at a given scale $m$.
When $m=1$, $GSM_m$ and $MFB_{m}$ are identical. However, when $m>1$, $MFB_{m}$ tends to
find more adequate testing signals than $GSM_m$, since the MODWPT zooms in the high frequency sub-bands by further decomposing $W_{m,n}$, while
the MODWT does not.

To use $MFB_{m}$ in practice, we need calculate $WV_{m,n}$ in (\ref{t2mv}) and replace $\Sigma_m$ in (\ref{jmf}) by a known matrix.
In general cases, $WV_{m,n}$ can be calculated by replacing $\sigma^2$ and ${\rm avar}(z_{m,n})$ with 
$\widehat{\sigma}^2$ and the NW estimator $\widehat{{\rm avar}}(z_{m,n})$,
and $\Sigma_{m}$ can be replaced by its NW estimator $\widehat{\Sigma}_{m}$, where
the $(i,j)$th entry of $\widehat{\Sigma}_{m}$ is
$$\widehat{\Sigma}_{m,i,j}=\frac{\widehat{{\rm acov}}(z_{m,i}z_{m,j})}{\sqrt{\widehat{{\rm avar}}(z_{m,i})}\sqrt{\widehat{{\rm avar}}(z_{m,j})}},$$
and $\widehat{{\rm acov}}(z_{m,i}z_{m,j})$ is the NW estimator of ${\rm acov}(z_{m,i}z_{m,j})$.
In a particular case, if $\{y_t\}$ satisfies the condition (\ref{cross_cumulant}), $WV_{m,n}$ can be calculated explicitly as in (\ref{pro1.1}), and
$\Sigma_{m}$ can be simplified as $A_{m}$ by the similar arguments as for Proposition \ref{p1}, where
the $(i,j)$th entry of $A_{m}$ is
\begin{equation} \label{dot_A}
A_{m,i,j}=\dfrac{a(\widetilde{v}_{m,i,j})}{\sqrt{a(\widetilde{v}_{m,i,i})}\sqrt{a(\widetilde{v}_{m,j,j})}}.
\end{equation}
Now, we consider three computational versions of
$MFB_{m}$:
\begin{itemize}
	\item $MFB_{m}^{g}$ calculates $WV_{m,n}$ as in (\ref{pro1.1}), and replaces $\Sigma_m$ by $A_m$ in (\ref{dot_A});

	\item $MFB_{m}^{\vartriangle}$ calculates $WV_{m,n}$  with $\sigma^2$ and ${\rm avar}(z_{m,n})$
replaced by $\widehat{\sigma}^2$ and  $\widehat{{\rm avar}}(z_{m,n})$,
and replaces $\Sigma_m$ by $A_m$;  

	\item $MFB_{m}^{e}$ calculates $WV_{m,n}$ with $\sigma^2$ and ${\rm avar}(z_{m,n})$ replaced by $\widehat{\sigma}^2$ and  $\widehat{{\rm avar}}(z_{m,n})$,
and replaces $\Sigma_m$ by $\widehat{\Sigma}_{m}$.
\end{itemize}
Note that $MFB_{m}^{g}$, $MFB_{m}^{\vartriangle}$ and $MFB_{m}^{e}$ are constructed in a similar
way as the multi-scale tests $GSM_{m}^{g}$, $GSM_{m}^{\vartriangle}$ and $GSM_{m}^{e}$ in Gen\c{c}ay and Signori (2015),
where we use the notation $GSM_{m}^{e}$ to denote their test $GSM_{m}$ for the notational consistency.
By construction, $MFB_{m}^{g}$ and $MFB_{m}^{\vartriangle}$ are feasible for the special case that
condition (\ref{cross_cumulant}) holds, while $MFB_{m}^{e}$ is valid for general cases.
The same conclusion holds for their multi-scale counterparts.

\section{Simulation}
In this section, we examine the finite-sample performance of our tests
$MFB_{m}^{g}$, $MFB_{m}^{\vartriangle}$ and $MFB_{m}^{e}$
in comparison with the portmanteau tests $Q_{K}$ in Ljung and Box (1978), the automatic portmanteau test $AQ$ in Escanciano and Lobato (2009), and the multi-scale tests $GSM_{m}^{g}$, $GSM_{m}^{\vartriangle}$ and $GSM_{m}^{e}$ in Gen\c{c}ay and Signori (2015).
Unless stated otherwise, all MFB and GSM tests are computed with Haar wavelet in the sequel.

\subsection{Size study}

Let $\epsilon_t\stackrel{i.i.d.}{\sim} N(0,1)$ unless specified.
To examine the empirical size of all tests, we
consider the following null models:
\begin{enumerate}[label=N\arabic*]
\item $[\bf{N(0,1)}]$ a standard normal process: $y_t=\epsilon_t$;
\item $[\bf{N(0,1)}$-$\bf{GARCH}]$ a GARCH process with $N(0,1)$ innovations:
	$y_t=\sigma_t\epsilon_t$ and $\sigma_t^2=0.001+0.05y^2_{t-1}+0.90\sigma^2_{t-1}$;
\item $[\bf{t_5}$-$\bf{GARCH}]$ a GARCH process as in model N2 except $\epsilon_t\stackrel{i.i.d.}{\sim} t_5$;
\item $[\bf{EGARCH}]$ an EGARCH process with $N(0,1)$ innovations:
	$y_t=\sigma_t\epsilon_t$ and $\log\sigma_t^2=0.001+0.5|\epsilon_t|-0.2\epsilon_t+0.95\log\sigma^2_{t-1}$;
\item $[\mbox{\bf{Mixture of normals}}]$ a mixture of two normals $N(0,1/2)$ and $N(0,1)$ with mixing probability $1/2$;
\item $[\bf{N(0,t)}]$: a heteroskedastic normal with trending variance: $y_t=\sqrt{t}\epsilon_t$;
\item $[\mbox{\bf{Time-varying GARCH}}]$ a time-varying GARCH$(1,1)$ process with $N(0,1)$ innovations:
	$y_t=\tau(t/T)u_t$, $\tau(x)=I(0<x<0.5)+2I(0.5\leq x<1)$,
    $u_t=\sigma_t\epsilon_t$ and $\sigma_t^2=0.05+0.05u^2_{t-1}+0.90\sigma^2_{t-1}$;
\item $[\mbox{\bf{Non-stationary GARCH}}]$ a non-stationary GARCH$(1,1)$ process with $N(0,1)$ innovations:
	$y_t=\sigma_t\epsilon_t$ and $\sigma_t^2=0.001+0.1096508y^2_{t-1}+0.90\sigma^2_{t-1}$;
\item $[\mbox{\bf{ZD-GARCH}}]$ a ZD-GARCH$(1,1)$ process with $N(0,1)$ innovations:
	$y_t=\sigma_t\epsilon_t$ and $\sigma_t^2=0.1096508y^2_{t-1}+0.90\sigma^2_{t-1}$;
\item $[\mbox{\bf{All-pass ARMA}}]$ an All-pass ARMA$(1,1)$ process with $N(0,1)$ innovations:
	$y_t=0.8y_{t-1}+\epsilon_t-(1/0.8)\epsilon_{t-1}$;
\item $[\mbox{\bf{Bilinear}}]$ a bilinear process with $N(0,1)$ innovations:
	$y_t=\epsilon_t+0.5\epsilon_{t-1}y_{t-2}$;
\item $[\mbox{\bf{Nonlinear MA}}]$ a nonlinear MA model with $N(0,1)$ innovations:
	$y_t=\epsilon_t+0.5\epsilon_{t-1}\epsilon_{t-2}$.
\end{enumerate}

Models N1--N6 were considered by Gen\c{c}ay and Signori (2015), and except model N6, the other five models
are stationary MDS with constant variances. Models N7--N9 were studied by Subba Rao (2006), Francq and Zako\"{i}an (2012), and
Li, Zhang, Zhu and Ling (2018), respectively. These three models are non-stationary MDS with time-varying variances.
Unlike models N1--N9, models N10--N12 are uncorrelated but non-MDS as shown in Shao (2011b).

\begin{table*}[htp!]
	\caption{Rejection rates (in percentage) under the null models N1--N12.}\label{reject}
\def\arraystretch{1.25}
	\setlength{\tabcolsep}{1.3mm}{
		\begin{tabular}{lcccccccccccccccc}
			\toprule[2pt]
			&&\multicolumn{3}{c}{$\mbox{N1}$}&&\multicolumn{3}{c}{$\mbox{N2}$}&&\multicolumn{3}{c}{$\mbox{N3}$}&&\multicolumn{3}{c}{$\mbox{N4}$}\\
            \cline{3-5} \cline{7-9} \cline{11-13} \cline{15-17}
            \multicolumn{1}{c}{$T$}&&100&300&1000&&100&300&1000&&100&300&1000&&100&300&1000\\
            \hline
            $MFB_{2}^{g}$&&4.56&4.82&4.62&&6.40&6.32&7.48&&7.20&9.68&11.88&&22.06&37.92&52.97\\
            $MFB_{2}^{\vartriangle}$&&9.23&6.08&5.37&&7.93&5.89&5.32&&7.49&5.78&5.36&&6.38&4.02&3.23\\
            $MFB_{2}^{e}$&&13.29&8.11&7.09&&11.86&7.43&6.67&&11.23&7.54&6.89&&10.88&6.92&4.70\\
			\cline{1-17}
	        $GSM_{2}^{g}$&&4.48&5.02&4.60&&5.90&5.96&7.04&&6.84&9.32&11.78&&18.98&33.34&46.88\\
            $GSM_{2}^{\vartriangle}$&&9.37&6.14&5.44&&8.41&6.30&5.39&&7.52&5.83&5.37&&6.39&3.92&3.19\\
            $GSM_{2}^{e}$&&13.42&8.20&7.23&&12.44&8.75&7.27&&11.29&7.60&6.95&&10.94&6.68&4.71\\
			$Q_5$&&5.54&4.98&4.94&&7.32&7.36&7.82&&8.90&10.82&14.96&&24.94&45.70&64.92\\
			$Q_{10}$&&6.04&5.16&5.10&&8.28&7.82&8.50&&9.26&12.50&16.48&&27.96&51.24&72.34\\
			$Q_{20}$&&7.68&5.74&5.92&&9.56&7.40&9.22&&9.92&12.22&17.16&&24.50&53.92&76.94\\
			AQ&&7.68&6.52&5.39&&7.71&6.35&5.93&&8.02&5.93&5.66&&6.68&5.75&5.39\\
			\midrule
			&&\multicolumn{3}{c}{$\mbox{N5}$}&&\multicolumn{3}{c}{$\mbox{N6}$}&&\multicolumn{3}{c}{$\mbox{N7}$}&&\multicolumn{3}{c}{$\mbox{N8}$}\\
            \cline{3-5} \cline{7-9} \cline{11-13} \cline{15-17}
            \multicolumn{1}{c}{$T$}&&100&300&1000&&100&300&1000&&100&300&1000&&100&300&1000\\
            \hline
            $MFB_{2}^{g}$&&4.34&5.02&5.00&&9.22&10.58&11.48&&11.43&14.40&16.70&&9.12&18.26&35.65\\
            $MFB_{2}^{\vartriangle}$&&9.39&6.34&5.32&&7.36&5.71&5.35&&7.09&5.58&5.03&&7.12&5.56&4.69\\
            $MFB_{2}^{e}$&&13.14&8.53&7.16&&10.97&7.68&6.54&&10.84&7.13&6.79&&11.24&7.69&5.97\\
			\cline{1-17}
	        $GSM_{2}^{g}$&&4.54&4.80&4.34&&8.98&10.02&10.72&&10.30&12.48&14.50&&7.96&16.54&33.27\\
            $GSM_{2}^{\vartriangle}$&&9.10&6.42&5.77&&7.65&5.87&5.39&&7.13&5.64&5.18&&7.06&5.57&4.82\\
            $GSM_{2}^{e}$&&13.11&8.60&7.23&&11.78&7.84&6.66&&11.08&7.19&6.84&&11.36&7.80&5.99\\
			$Q_5$&&5.54&5.44&5.00&&12.48&13.84&13.82&&14.52&19.25&19.01&&10.10&22.44&50.18\\
			$Q_{10}$&&6.06&5.24&4.70&&16.30&17.14&17.84&&19.38&24.60&27.64&&12.14&29.06&63.70\\
			$Q_{20}$&&7.24&5.64&5.76&&19.96&24.82&14.62&&22.03&31.05&36.87&&13.68&34.90&76.68\\
			AQ&&7.38&6.72&5.50&&7.90&6.40&5.42&&7.43&6.23&5.91&&7.26&6.64&5.68\\
			\midrule
			&&\multicolumn{3}{c}{$\mbox{N9}$}&&\multicolumn{3}{c}{$\mbox{N10}$}&&\multicolumn{3}{c}{$\mbox{N11}$}&&\multicolumn{3}{c}{$\mbox{N12}$}\\
            \cline{3-5} \cline{7-9} \cline{11-13} \cline{15-17}
            \multicolumn{1}{c}{$T$}&&100&300&1000&&100&300&1000&&100&300&1000&&100&300&1000\\
            \hline
            $MFB_{2}^{g}$&&9.18&19.20&37.75&&5.02&5.08&4.94&&12.90&16.24&18.84&&7.78&8.86&10.24\\
            $MFB_{2}^{\vartriangle}$&&6.95&5.38&4.62&&8.32&6.58&5.29&&7.11&5.53&5.09&&7.59&5.93&5.49\\
            $MFB_{2}^{e}$&&11.26&7.69&5.83&&12.76&8.07&7.48&&11.06&7.09&6.40&&11.35&7.67&6.42\\
			\cline{1-17}
	        $GSM_{2}^{g}$&&8.06&17.72&35.55&&5.64&5.54&5.53&&11.98&14.58&17.30&&7.86&9.50&10.56\\
            $GSM_{2}^{\vartriangle}$&&7.01&5.41&4.68&&8.29&6.73&5.33&&7.09&5.55&5.16&&7.67&6.03&5.64\\
            $GSM_{2}^{e}$&&11.29&7.76&5.85&&12.89&8.11&7.54&&11.03&7.16&6.42&&11.70&7.82&6.55\\
			$Q_5$&&10.18&22.82&50.62&&5.32&5.08&5.48&&13.94&15.62&16.35&&7.84&8.92&9.85\\
			$Q_{10}$&&12.28&29.52&63.98&&6.42&5.26&5.15&&11.30&12.82&13.58&&8.36&8.14&8.03\\
			$Q_{20}$&&13.88&35.38&77.08&&7.10&5.66&5.20&&11.48&10.88&9.85&&9.38&7.82&6.74\\
			AQ&&7.10&6.62&5.73&&8.84&7.70&6.43&&9.96&9.02&8.82&&8.08&6.74&6.20\\
			\bottomrule[2pt]
		\end{tabular}
	}
\end{table*}

As the settings in Gen\c{c}ay and Signori (2015),
Table \ref{reject} reports the proportion (in percentage) of rejections at 5\%
nominal level for all MFB and GSM tests with $m=2$, the portmanteau tests $Q_{K}$ with $K=5, 10, 20$, and the automatic portmanteau test $AQ$,
where 10000 replications are generated from each null model with the sample size $T=100$, 300 or 1000.
From this table, our findings are as follows:

(i) Our three MFB tests have a similar size performance as their GSM counterparts in all examined cases.
When the sample size is small (e.g., $T=100$), $MFB_2^g$ has an accurate size performance, except for models N4, N6--N9 and N11--N12.
As the sample size becomes larger (e.g., $T=1000$), the over-sized problem for $MFB_2^g$ is even worse.
In contrast, $MFB_{2}^{\vartriangle}$ and $MFB_{2}^{e}$ can always have accurate sizes when the sample size is large, although
they (particularly $MFB_{2}^{e}$) tend to be slightly over-sized when the sample size is small.

(ii) All three portmanteau tests $Q_{K}$ show good size performances in models N1, N5 and N10, but they
have the severe over-sized problem in models N3--N4, N6--N9 and N11, and this problem tends to exist in models N2 and N12
even when the sample size is large (e.g., $T=1000$).

(iii) The automatic portmanteau test $AQ$ exhibits a good size performance in all examined cases, except that it
tends to have a slightly over-sized problem when the sample size is small, and this problem remains in models N10--N12 even when the
sample size is large.

Overall, our findings are similar to those in Gen\c{c}ay and Signori (2015). On one hand, when the sample size is small, $MFB_2^g$ (or $GSM_2^g$) has a relatively better size performance than others for most of stationary MDS data, and
$MFB_2^{\vartriangle}$ (or $GSM_2^{\vartriangle}$ and $AQ$) does this for most of non-stationary or non-MDS data.
On the other hand, when the sample size is large, $MFB_{2}^{\vartriangle}$ (or $GSM_{2}^{\vartriangle}$) seems to have the best
size performance in general.

\subsection{Power study}

To examine the empirical power of all tests, we consider the following four alternative models:

\begin{enumerate}[label=A\arabic*]
\item $[\bf{N(0,1)}$-$\bf{AR(2)}]$ an AR(2) process with $N(0,1)$ innovations: $y_t=\beta_1y_{t-1}+\beta_2y_{t-2}+\epsilon_t$;
\item $[\bf{N(0,1)}$-$\bf{AR(3)}]$ an AR(3) process with $N(0,1)$ innovations: $y_t=\beta_1y_{t-1}+\beta_2y_{t-3}+\epsilon_t$;
\item $[\bf{N(0,t)}$-$\bf{AR(2)}]$ an AR(2) process with $N(0,t)$ innovations: $y_t=\beta_1y_{t-1}+\beta_2y_{t-2}+\sqrt{t}\epsilon_t$;
\item $[\bf{N(0,t)}$-$\bf{AR(3)}]$ an AR(3) process with $N(0,t)$ innovations: $y_t=\beta_1y_{t-1}+\beta_2y_{t-3}+\sqrt{t}\epsilon_t$,
\end{enumerate}
where $\beta_1$ (or $\beta_2$) is set to be $-0.30, -0.20, ..., 0.20,$ and $0.30$.

Model A1 was considered in Gen\c{c}ay and Signori (2015), and models A2--A4 are designed to see how the tests perform when the data have the serial dependence at a larger lag or they are heteroskedastic.

As before, we follow the settings in Gen\c{c}ay and Signori (2015), and thus restrict our analysis to compare the
(size-adjusted) power of $MFB_2^g$, $GSM_2^g$, $Q_{20}$, and $AQ$ when the sample size is small.
Tables \ref{sp1} and \ref{sp2} report the power (in percentage) at 5\%
nominal level for $MFB_2^g$, where 10000 replications are generated from each alternative model with the sample size $T=100$.
To make a comparison, Tables \ref{sp1} and \ref{sp2} also report the relative power gains of $MFB_2^g$ with respect to the other three tests.
From these two tables, we can have the following findings:

(i) For model A1, $MFB_2^g$ is generally more powerful than $GSM_2^g$ when $\beta_1<0$,
while  $GSM_2^g$ outperforms $MFB_2^g$ when $\beta_1>0$. For model A2, the advantage of $GSM_2^g$ over $MFB_2^g$ largely disappears, but
$MFB_2^g$ has a huge power improvement over $GSM_2^g$ up to 786\%. This implies that the power advantage of $MFB_2^g$ over
$GSM_2^g$ tends to be more substantial, when the serial dependence of data happens at larger lags. For models A3--A4 with heteroskedastic data, a similar conclusion can be drawn.

(ii) For all considered four models, $MFB_2^g$ is always more powerful than $Q_{20}$. The power performance between
$MFB_2^g$ and $AQ$ is mixed. For models A1 and A3, $MFB_2^g$ (or $AQ$)  shows its relative better performance when $\beta_1>0$ (or $\beta_1<0$).
For model A2, $MFB_2^g$ has a clear power improvement over $AQ$ up to 88\%, while $AQ$ is only slightly better than $MFB_2^g$ when $\beta_1<0$ and $\beta_2$ is close to 0. For model A4, a similar phenomenon as for model A2 can be observed. All these findings once again imply that
$MFB_2^g$ has a more substantial power advantage over $AQ$, when the serial dependence of data happens at larger lags.

\begin{table*}[htp!]
	\caption{Size-adjusted power and relative power against model A1 (left side) and model A2 (right side).}\label{sp1}
	\def\arraystretch{1.3}
	\setlength{\tabcolsep}{0.7mm}{
		\begin{tabular}{ccccccccccccccccc}
			\toprule[2pt]
			\multicolumn{8}{c}{$A1: y_t=\beta_1y_{t-1}+\beta_2y_{t-2}+\epsilon_t$}&&\multicolumn{8}{c}{$A2: y_t=\beta_1y_{t-1}+\beta_2y_{t-3}+\epsilon_t$}\\
			\cline{1-8}\cline{10-17}
			&\multicolumn{7}{c}{$MFB_2^{g}$}&&\ &\multicolumn{7}{c}{$MFB_2^{g}$}\\
			\cline{2-8}\cline{11-17}
			 \diagbox[width=3em]{$\beta_1$}{$\beta_2$}&\underline{0.30}&\underline{0.20}&\underline{0.10}&\underline{0.00}&\underline{-0.10}&\underline{-0.20}&\underline{-0.30}&\ &\diagbox[width=3em]{$\beta_1$}{$\beta_2$}&\underline{0.30}&\underline{0.20}&\underline{0.10}&\underline{0.00}&\underline{-0.10}&\underline{-0.20}&\underline{-0.30}\\
			\underline{0.30}&98.23&93.83&82.47&72.27&63.47&72.50&86.20&\ &\underline{0.30}&95.07&87.10&78.03&71.27&76.37&86.77&95.27\\
			\underline{0.20}&91.93&76.23&55.27&34.87&33.57&50.10&77.13&\ &\underline{0.20}&86.27&65.90&44.57&37.70&46.70&66.90&87.37\\
			\underline{0.10}&79.27&49.77&25.50&11.70&14.47&36.17&70.10&\ &\underline{0.10}&74.97&44.13&20.90&13.67&22.73&44.03&74.60\\
			\underline{0.00}&74.13&38.63&14.53&5.16&10.47&33.30&67.17&\ &\underline{0.00}&70.13&34.43&12.13&5.17&11.67&36.50&68.90\\
			\underline{-0.10}&80.83&50.67&23.70&12.30&13.83&35.90&70.90&\ &\underline{-0.10}&75.67&41.67&16.07&9.70&16.50&43.17&75.07\\
			\underline{-0.20}&92.37&76.33&53.20&36.57&32.07&49.97&77.43&\ &\underline{-0.20}&85.37&64.07&40.13&32.20&41.37&63.30&85.30\\
			\underline{-0.30}&98.33&92.87&84.03&71.60&64.97&71.73&87.10&\ &\underline{-0.30}&95.00&84.80&72.43&67.93&73.77&84.80&95.33\\
			\cline{1-8}\cline{10-17}
			
			&\multicolumn{7}{c}{Relative power: $(MFB_2^{g}/GSM_{2}^{g})-1$}&&\ &\multicolumn{7}{c}{Relative power: $(MFB_2^{g}/GSM_{2}^{g})-1$}\\
			\cline{2-8}\cline{11-17}
			 \diagbox[width=3em]{$\beta_1$}{$\beta_2$}&\underline{0.30}&\underline{0.20}&\underline{0.10}&\underline{0.00}&\underline{-0.10}&\underline{-0.20}&\underline{-0.30}&\ &\diagbox[width=3em]{$\beta_1$}{$\beta_2$}&\underline{0.30}&\underline{0.20}&\underline{0.10}&\underline{0.00}&\underline{-0.10}&\underline{-0.20}&\underline{-0.30}\\
			\underline{0.30}&0.00&-$\bm{0.01}$&-$\bm{0.04}$&-$\bm{0.08}$&-$\bm{0.12}$&-$\bm{0.09}$&-$\bm{0.05}$&\ &\underline{0.30}&0.00&-\textbf{0.03}&-\textbf{0.04}&0.05&0.37&0.89&1.09\\
			\underline{0.20}&0.00&-\textbf{0.03}&-\textbf{0.10}&-\textbf{0.17}&-\textbf{0.19}&-\textbf{0.13}&-\textbf{0.05}&\ &\underline{0.20}&0.05&-\textbf{0.03}&-\textbf{0.11}&0.03&0.85&2.43&3.05\\
			\underline{0.10}&0.00&-\textbf{0.04}&-\textbf{0.12}&-\textbf{0.18}&-\textbf{0.18}&-\textbf{0.08}&0.00&\ &\underline{0.10}&0.48&0.22&-\textbf{0.10}&0.00&1.52&4.18&5.41\\
			\underline{0.00}&0.05&0.05&-\textbf{0.01}&0.00&0.05&0.10&0.10&\ &\underline{0.00}&2.19&1.62&0.75&0.01&0.68&2.05&2.12\\
			\underline{-0.10}&0.07&0.09&0.11&0.05&0.17&0.31&0.20&\ &\underline{-0.10}&7.86&7.68&2.41&-\textbf{0.09}&-\textbf{0.07}&0.28&0.49\\
			\underline{-0.20}&0.04&0.04&0.01&-\textbf{0.06}&0.03&0.24&0.20&\ &\underline{-0.20}&8.23&6.45&1.45&0.07&-\textbf{0.12}&-\textbf{0.04}&0.07\\
			\underline{-0.30}&0.00&0.01&-\textbf{0.01}&-\textbf{0.04}&-\textbf{0.05}&0.05&0.11&\ &\underline{-0.30}& 3.27&1.78&0.66&0.06&-\textbf{0.05}&-\textbf{0.05}&0.00\\
			\cline{1-8}\cline{10-17}
			
			&\multicolumn{7}{c}{Relative power: $(MFB_2^{g}/Q_{20})-1$}&&\ &\multicolumn{7}{c}{Relative power: $(MFB_2^{g}/Q_{20})-1$}\\
			\cline{2-8}\cline{11-17}
			 \diagbox[width=3em]{$\beta_1$}{$\beta_2$}&\underline{0.30}&\underline{0.20}&\underline{0.10}&\underline{0.00}&\underline{-0.10}&\underline{-0.20}&\underline{-0.30}&\ &\diagbox[width=3em]{$\beta_1$}{$\beta_2$}&\underline{0.30}&\underline{0.20}&\underline{0.10}&\underline{0.00}&\underline{-0.10}&\underline{-0.20}&\underline{-0.30}\\
			\underline{0.30}&0.13&0.34&0.63&1.02&1.43&1.14&0.79&\ &\underline{0.30}&0.27&0.57&0.98&0.85&0.70&0.30&0.04\\
			\underline{0.20}&0.46&0.87&1.13&1.29&1.09&1.11&0.74&\ &\underline{0.20}&0.61&1.09&1.37&1.27&0.93&0.66&0.34\\
			\underline{0.10}&0.80&1.25&1.06&0.70&0.54&0.95&0.81&\ &\underline{0.10}&1.03&1.52&1.40&0.79&0.97&0.82&0.62\\
			\underline{0.00}&1.00&1.35&0.62&-\textbf{0.02}&0.31&1.02&0.80&\ &\underline{0.00}&1.06&1.31&0.86&0.02&0.61&1.03&0.75\\
			\underline{-0.10}&0.67&0.82&0.86&0.47&0.45&0.87&0.85&\ &\underline{-0.10}&0.74&1.00&0.65&0.22&0.81&0.96&0.77\\
			\underline{-0.20}&0.28&0.58&0.78&0.87&0.93&1.23&0.78&\ &\underline{-0.20}&0.36&0.80&0.93&0.94&0.91&0.65&0.30\\
			\underline{-0.30}&0.08&0.23&0.45&0.83&1.16&1.13&0.76&\ &\underline{-0.30}&0.09&0.34&0.60&0.85&0.57&0.37&0.25\\
			\cline{1-8}\cline{10-17}
			
			&\multicolumn{7}{c}{Relative power: $(MFB_2^{g}/AQ)-1$}&&\ &\multicolumn{7}{c}{Relative power: $(MFB_2^{g}/AQ)-1$}\\
			\cline{2-8}\cline{11-17}
			 \diagbox[width=3em]{$\beta_1$}{$\beta_2$}&\underline{0.30}&\underline{0.20}&\underline{0.10}&\underline{0.00}&\underline{-0.10}&\underline{-0.20}&\underline{-0.30}&\ &\diagbox[width=3em]{$\beta_1$}{$\beta_2$}&\underline{0.30}&\underline{0.20}&\underline{0.10}&\underline{0.00}&\underline{-0.10}&\underline{-0.20}&\underline{-0.30}\\
			\underline{0.30}&0.02&0.06&0.07&0.03&0.03&0.08&0.06&\ &\underline{0.30}&0.09&0.12&0.05&0.00&0.00&0.05&0.04\\
			\underline{0.20}&0.04&0.14&0.18&0.07&0.01&0.02&0.00&\ &\underline{0.20}&0.26&0.39&0.16&0.01&0.08&0.17&0.12\\
			\underline{0.10}&0.02&0.12&0.31&0.14&-\textbf{0.12}&-\textbf{0.15}&-\textbf{0.08}&\ &\underline{0.10}&0.48&0.85&0.63&0.13&0.38&0.37&0.24\\
			\underline{0.00}&0.00&0.02&0.07&0.03&-\textbf{0.28}&-\textbf{0.23}&-\textbf{0.12}&\ &\underline{0.00}&0.56&0.88&0.56&0.00&0.48&0.57&0.36\\
			\underline{-0.10}&0.01&0.02&-\textbf{0.01}&-\textbf{0.12}&-\textbf{0.20}&-\textbf{0.16}&-\textbf{0.08}&\ &\underline{-0.10}&0.30&0.43&0.09&-\textbf{0.31}&0.07&0.42&0.26\\
			\underline{-0.20}&0.01&0.04&-\textbf{0.01}&-\textbf{0.11}&-\textbf{0.12}&-\textbf{0.03}&0.00&\ &\underline{-0.20}&0.10&0.08&-\textbf{0.10}&-\textbf{0.21}&-\textbf{0.09}&0.06&0.09\\
			\underline{-0.30}&0.01&0.00&-\textbf{0.01}&-\textbf{0.05}&-\textbf{0.06}&0.00&0.05&\ &\underline{-0.30}&0.02&0.01&-\textbf{0.07}&-\textbf{0.11}&-\textbf{0.07}&0.01&0.02\\		
			\bottomrule[2pt]
		\end{tabular}
\begin{tablenotes}
      \item[1] {Note: The value of relative power less than zero is in boldface.}
  \end{tablenotes}
	}
\end{table*}

\begin{table*}[htp!]
	\caption{Size-adjusted power and relative power against model A3 (left side) and model A4 (right side).}\label{sp2}
\def\arraystretch{1.3}
	\setlength{\tabcolsep}{0.7mm}{
		\begin{tabular}{ccccccccccccccccc}
			\toprule[2pt]
\multicolumn{8}{c}{$A3: y_t=\beta_1y_{t-1}+\beta_2y_{t-2}+\sqrt{t}\epsilon_t$}&&\multicolumn{8}{c}{$A4: y_t=\beta_1y_{t-1}+\beta_2y_{t-3}+\sqrt{t}\epsilon_t$}\\
\cline{1-8}\cline{10-17}
			&\multicolumn{7}{c}{$MFB_2^{g}$}&&\ &\multicolumn{7}{c}{$MFB_2^{g}$}\\
            \cline{2-8}\cline{11-17}
			 \diagbox[width=3em]{$\beta_1$}{$\beta_2$}&\underline{0.30}&\underline{0.20}&\underline{0.10}&\underline{0.00}&\underline{-0.10}&\underline{-0.20}&\underline{-0.30}&\ &\diagbox[width=3em]{$\beta_1$}{$\beta_2$}&\underline{0.30}&\underline{0.20}&\underline{0.10}&\underline{0.00}&\underline{-0.10}&\underline{-0.20}&\underline{-0.30}\\
			\underline{0.30}&95.53&87.08&74.23&58.87&50.30&57.11&73.14&\ &\underline{0.30}&89.75&78.22&64.65&59.69&61.71&72.84&86.95\\
			\underline{0.20}&84.13&65.73&43.71&29.02&24.44&36.81&62.72&\ &\underline{0.20}&75.44&53.40&35.79&27.89&31.26&49.21&72.57\\
			\underline{0.10}&67.94&41.10&19.75&9.85&11.49&27.23&55.84&\ &\underline{0.10}&60.54&33.04&15.83&9.58&14.77&30.94&59.84\\
			\underline{0.00}&60.15&29.84&11.02&4.91&8.29&24.26&53.07&\ &\underline{0.00}&55.02&25.90&9.10&4.80&9.63&26.76&55.12\\
			\underline{-0.10}&67.94&40.41&19.44&9.85&11.27&27.09&55.15&\ &\underline{-0.10}&59.89&32.05&14.77&10.15&16.71&33.87&61.17\\
			\underline{-0.20}&84.54&65.50&43.57&28.17&24.66&36.94&62.66&\ &\underline{-0.20}&72.95&49.16&32.13&28.58&35.67&54.12&75.28\\
			\underline{-0.30}&95.53&87.36&72.95&58.58&51.01&56.96&73.69&\ &\underline{-0.30}&86.90&73.15&61.55&58.96&65.48&77.84&89.66\\
			\cline{1-8}\cline{10-17}

			&\multicolumn{7}{c}{Relative power: $(MFB_2^{g}/GSM_{2}^{g})-1$}&&\ &\multicolumn{7}{c}{Relative power: $(MFB_2^{g}/GSM_{2}^{g})-1$}\\
            \cline{2-8}\cline{11-17}
			 \diagbox[width=3em]{$\beta_1$}{$\beta_2$}&\underline{0.30}&\underline{0.20}&\underline{0.10}&\underline{0.00}&\underline{-0.10}&\underline{-0.20}&\underline{-0.30}&\ &\diagbox[width=3em]{$\beta_1$}{$\beta_2$}&\underline{0.30}&\underline{0.20}&\underline{0.10}&\underline{0.00}&\underline{-0.10}&\underline{-0.20}&\underline{-0.30}\\
			\underline{0.30}&0.00&0.01&-\textbf{0.07}&-\textbf{0.12}&-\textbf{0.18}&-\textbf{0.15}&-\textbf{0.07}&\ &\underline{0.30}&0.06&0.00&-\textbf{0.07}&-\textbf{0.11}&-\textbf{0.07}&0.06&0.16\\
			\underline{0.20}&0.02&-\textbf{0.07}&-\textbf{0.14}&-\textbf{0.20}&-\textbf{0.24}&-\textbf{0.21}&-\textbf{0.05}&\ &\underline{0.20}&0.26&0.09&-\textbf{0.10}&-\textbf{0.17}&-\textbf{0.06}&0.22&0.46\\
			\underline{0.10}&0.00&-\textbf{0.07}&-\textbf{0.19}&-\textbf{0.21}&-\textbf{0.20}&-\textbf{0.14}&0.00&\ &\underline{0.10}&0.77&0.45&-\textbf{0.02}&-\textbf{0.17}&0.07&0.70&1.05\\
			\underline{0.00}&0.05&0.03&-\textbf{0.03}&-\textbf{0.02}&0.00&0.10&0.09&\ &\underline{0.00}&2.22&1.48&0.42&-\textbf{0.03}&0.72&1.44&2.21\\
			\underline{-0.10}&0.08&0.12&0.06&-\textbf{0.02}&0.23&0.40&0.28&\ &\underline{-0.10}&2.34&1.50&0.45&-\textbf{0.02}&0.49&1.29&2.00\\
			\underline{-0.20}&0.05&0.06&0.01&-\textbf{0.04}&0.08&0.28&0.35&\ &\underline{-0.20}&0.80&0.44&0.03&-\textbf{0.06}&0.17&0.51&0.83\\
			\underline{-0.30}&0.02&0.01&-\textbf{0.02}&-\textbf{0.05}&-\textbf{0.06}&0.09&0.20&\ &\underline{-0.30}&0.19&0.09&-\textbf{0.01}&-\textbf{0.04}&0.01&0.13&0.20\\
            \cline{1-8}\cline{10-17}

&\multicolumn{7}{c}{Relative power: $(MFB_2^{g}/Q_{20})-1$}&&\ &\multicolumn{7}{c}{Relative power: $(MFB_2^{g}/Q_{20})-1$}\\
            \cline{2-8}\cline{11-17}
			 \diagbox[width=3em]{$\beta_1$}{$\beta_2$}&\underline{0.30}&\underline{0.20}&\underline{0.10}&\underline{0.00}&\underline{-0.10}&\underline{-0.20}&\underline{-0.30}&\ &\diagbox[width=3em]{$\beta_1$}{$\beta_2$}&\underline{0.30}&\underline{0.20}&\underline{0.10}&\underline{0.00}&\underline{-0.10}&\underline{-0.20}&\underline{-0.30}\\
			\underline{0.30}&0.24&0.50&0.85&1.14&1.31&1.26&0.94&\ &\underline{0.30}&0.45&0.70&0.93&1.18&1.07&0.81&0.48\\
			\underline{0.20}&0.62&0.95&1.11&1.06&1.07&1.00&0.90&\ &\underline{0.20}&0.77&1.05&1.34&1.27&1.02&0.87&0.65\\
			\underline{0.10}&0.96&1.27&1.10&0.56&0.48&0.82&0.92&\ &\underline{0.10}&1.14&1.27&0.82&0.45&0.57&0.84&0.78\\
			\underline{0.00}&1.17&1.30&0.60&-\textbf{0.02}&0.11&0.75&0.86&\ &\underline{0.00}&1.20&1.13&0.56&0.00&0.30&0.83&0.84\\
			\underline{-0.10}&0.72&0.91&0.74&0.39&0.37&0.74&0.91&\ &\underline{-0.10}&0.89&0.94&0.67&0.50&0.69&0.79&0.74\\
			\underline{-0.20}&0.42&0.65&0.83&0.85&0.92&0.97&0.92&\ &\underline{-0.20}&0.70&0.89&0.99&0.94&0.80&0.72&0.50\\
			\underline{-0.30}&0.14&0.32&0.57&0.95&1.16&1.16&0.94&\ &\underline{-0.30}&0.46&0.74&1.00&0.87&0.64&0.43&0.24\\
            \cline{1-8}\cline{10-17}

&\multicolumn{7}{c}{Relative power: $(MFB_2^{g}/AQ)-1$}&&\ &\multicolumn{7}{c}{Relative power: $(MFB_2^{g}/AQ)-1$}\\
            \cline{2-8}\cline{11-17}
			 \diagbox[width=3em]{$\beta_1$}{$\beta_2$}&\underline{0.30}&\underline{0.20}&\underline{0.10}&\underline{0.00}&\underline{-0.10}&\underline{-0.20}&\underline{-0.30}&\ &\diagbox[width=3em]{$\beta_1$}{$\beta_2$}&\underline{0.30}&\underline{0.20}&\underline{0.10}&\underline{0.00}&\underline{-0.10}&\underline{-0.20}&\underline{-0.30}\\
			\underline{0.30}&0.03&0.06&0.05&0.02&0.04&0.09&0.07&\ &\underline{0.30}&0.08&0.10&0.05&0.00&0.00&0.05&0.03\\
			\underline{0.20}&0.04&0.11&0.17&0.08&0.01&0.03&0.01&\ &\underline{0.20}&0.22&0.34&0.16&0.00&0.07&0.15&0.11\\
			\underline{0.10}&0.01&0.13&0.28&0.13&-\textbf{0.09}&-\textbf{0.15}&-\textbf{0.09}&\ &\underline{0.10}&0.45&0.82&0.56&0.12&0.36&0.33&0.20\\
			\underline{0.00}&0.01&0.03&0.06&0.03&-\textbf{0.25}&-\textbf{0.21}&-\textbf{0.10}&\ &\underline{0.00}&0.50&0.84&0.51&-\textbf{0.01}&0.43&0.57&0.34\\
			\underline{-0.10}&0.02&0.03&-\textbf{0.02}&-\textbf{0.10}&-\textbf{0.19}&-\textbf{0.16}&-\textbf{0.07}&\ &\underline{-0.10}&0.28&0.40&0.07&-\textbf{0.32}&0.07&0.39&0.25\\
			\underline{-0.20}&0.01&0.05&-\textbf{0.01}&-\textbf{0.10}&-\textbf{0.11}&-\textbf{0.02}&0.01&\ &\underline{-0.20}&0.10&0.07&-\textbf{0.10}&-\textbf{0.22}&-\textbf{0.10}&0.05&0.08\\
			\underline{-0.30}&0.02&0.00&0.00&-\textbf{0.06}&-\textbf{0.05}&0.01&0.06&\ &\underline{-0.30}&0.02&0.00&-\textbf{0.06}&-\textbf{0.10}&-\textbf{0.07}&0.02&0.02\\		
			\bottomrule[2pt]
		\end{tabular}
\begin{tablenotes}
      \item[1] {Note: The value of relative power less than zero is in boldface.}
  \end{tablenotes}
	}
\end{table*}

\subsection{Robust analysis}

\begin{figure}[htp!]
	\centering
    \includegraphics[width=6in,height=3in]{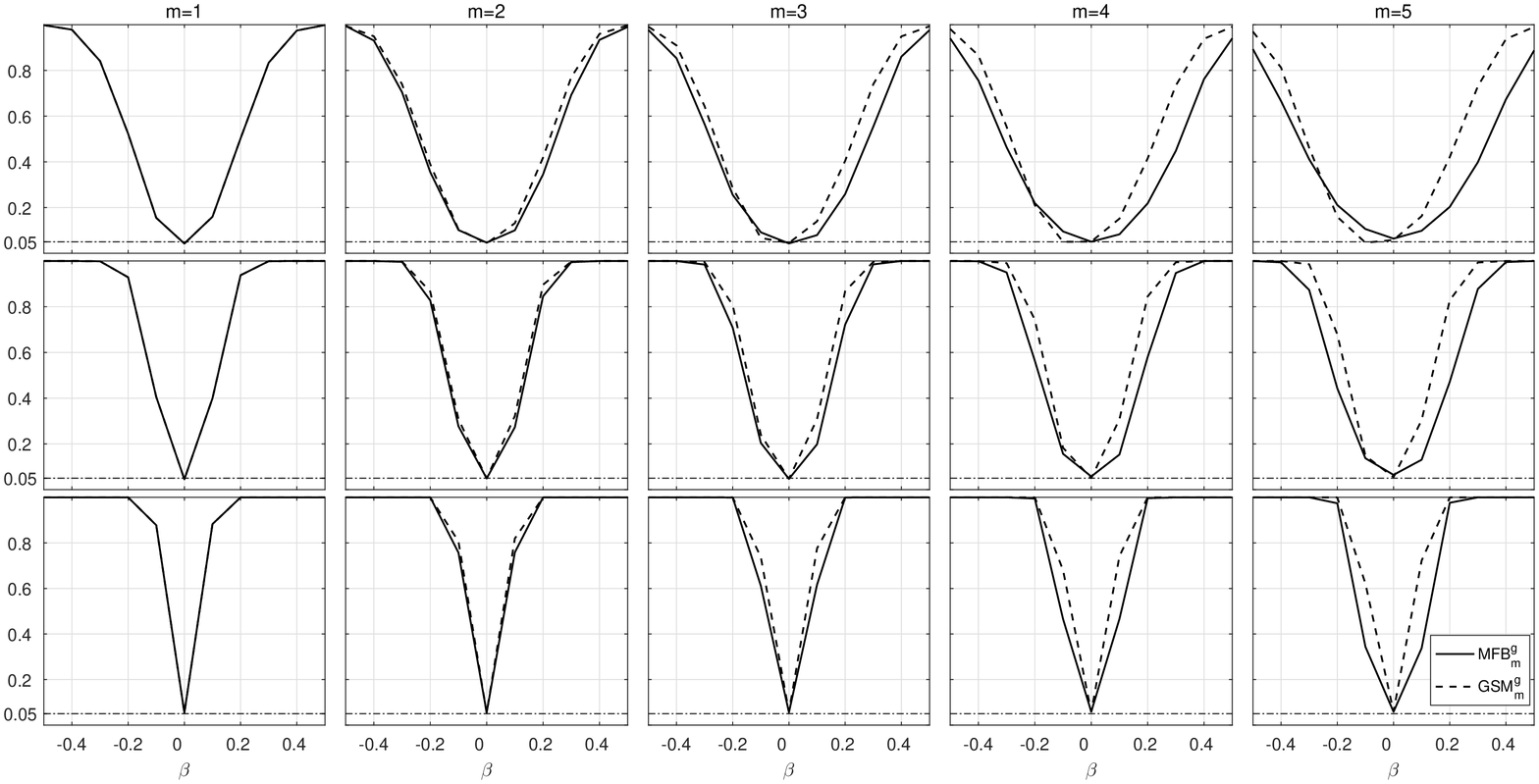}
	\caption{The power of $MFB_m^g$ and $GSM_m^g$ (for $m=1,...,5$) against $AR(1)$ alternative: $y_t=\beta y_{t-1}+\epsilon_t$.
The top, middle and bottom panels are corresponding to the sample size $T=100, 300,$ and 1000, respectively.}\label{AR1}
\end{figure}

\begin{figure}[htp!]
	\centering
	\includegraphics[width=6in,height=3in]{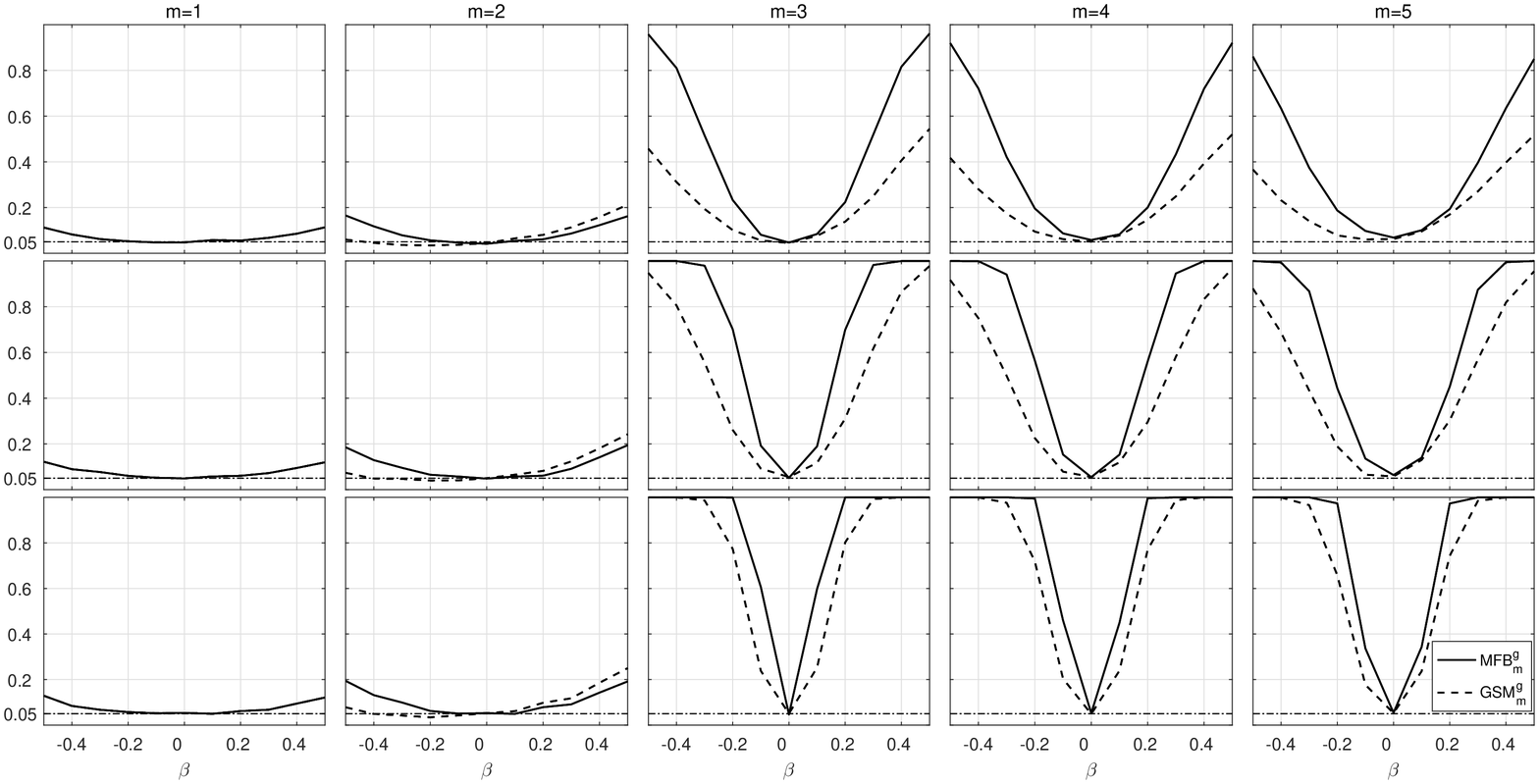}
	\caption{The power of $MFB_m^g$ and $GSM_m^g$ (for $m=1,...,5$) against $AR(5)$ alternative: $y_t=\beta y_{t-5}+\epsilon_t$.
The top, middle and bottom panels are corresponding to the sample size $T=100, 300,$ and 1000, respectively.}\label{AR5}
\end{figure}

In the previous two subsections, we focus on $m=2$ for our MFB tests. This subsection aims to do some robust analysis for our
MFB tests, based on the settings as in Gen\c{c}ay and Signori (2015). First, we explore the finite sample performance of our MFB tests in terms of the choice of $m$. To illustrate it,
we generate $10000$ replications with sample size $T=100, 300$ or 1000 from the following AR($k$) model:
$$y_t=\beta y_{t-k}+\epsilon_t,$$
where $|\beta|<1$. Figures\,\ref{AR1} and \ref{AR5} plot the (size-adjusted) power of $MFB_m^g$ (for $m=1,...,5$) against AR($1$) and AR($5$) models at 5\% nominal
level, respectively. As a comparison, the (size-adjusted) power of $GSM_m^g$
is also plotted in these two figures. From Figure\,\ref{AR1}, we can find that when $m=1, 2$ and $3$, all MFB and GSM tests have similar power performances, and when
$m=4$ and $5$, the GSM tests perform better than the MFB tests especially for $\beta>0$. In contrast, Figure\,\ref{AR5} shows that
when $m=3, 4$ and $5$, the MFB tests are clearly more powerful than the GSM tests, while all tests exhibit low power when $m=1$ and $2$.
These findings suggest that when the serial dependence happens at the small lag, our MFB tests can perform stably over $m$, and when
the serial dependence happens at the large lag, our MFB tests with a large $m$ can perform well, and they are generally more powerful than
the GSM tests in this case.

%

Second, we check the finite sample performance of our MFB tests in terms of the choice of wavelets.
As the settings in Gen\c{c}ay and Signori (2015), we report
the size and (size-adjusted) power of $MFB_2^g$ for Haar wavelet and Daubechies wavelets D(4), D(6), D(8) and D(10) in Table \ref{wavelet}.
From this table, we can see that there is no significant difference in terms of size, but the Haar wavelet has some marginal advantages in terms of power.

\begin{table*}[h!]
	\caption{Size and power of $MFB_2^g$ for various wavelets.}\label{wavelet}
	\setlength{\tabcolsep}{1mm}{
		\begin{tabular}{ccccccc}
			\toprule[2pt]
			Models&$T$&Haar&D(4)&D(6)&D(8)&D(10)\\
			\midrule
			\multicolumn{7}{l}{Panel A: size study}\\
			\midrule
			\multirow{3}{*}{Model N1}&100&4.65&4.55&4.55&4.54&4.60\\
			&300&4.68&4.69&4.75&4.70&4.65\\
			&1000&4.42&4.58&4.59&4.52&4.48\\
			\midrule
			\multirow{3}{*}{Model N2}&100&5.14&5.40&5.26&5.19&5.26\\
			&300&6.69&6.62&6.65&6.58&6.59\\
			&1000&7.22&7.10&7.05&7.06&7.08\\
			\midrule
			\multicolumn{7}{l}{Panel B: power study}\\
			\midrule
			\multirow{3}{*}{\makecell{Model A1\\with $\beta_1=\beta_2=0.1$}}&100&19.43&17.65&16.39&15.84&15.20\\
			&300&51.83&48.55&46.23&44.59&43.40\\
			&1000&97.68&97.07&96.34&95.81&95.34\\
			\midrule
			\multirow{3}{*}{\makecell{Model A3\\with $\beta_1=\beta_2=0.1$}}&100&24.46&23.25&22.37&21.99&21.63\\
			&300&54.89&52.40&50.93&49.91&49.15\\
			&1000&96.15&95.49&94.77&94.23&93.92\\
			\bottomrule[2pt]
		\end{tabular}
	}
\end{table*}

\section{Applications}

\subsection{Application 1}
Checking whether the market index returns are predictable has been a long standing problem
in the literature. The empirical studies in Lo and MacKinlay (1988) and Hong and Lee (2005) found that the S\&P500 index returns are predictable.
However, their empirical studies overlooked a fact that
the predictability conclusion made based on the entire period may not be true for some specific sub-periods. To relieve this concern,
we examine whether the recent S\&P500 return series as well as their sub-series are white noises, and if the white noise assumption is rejected, the examined series is predictable, therefore giving the empirical evidence against the efficient market hypothesis.

We consider the daily S\&P500 index from January 2, 2006 to December 31, 2015, with 2515 observations in total.
Denote the S\&P500 return $y_t=100\log(P_t/P_{t-1})$, where $P_t$ is the closing S\&P500 index at day $t$.
We first apply the MFB tests, the GSM tests and the AQ test to the entire 10-year return series, and the results in Panel A of
Table \ref{sp500} show a very strong evidence to reject the white noise assumption for this entire series.
Although the entire series is not  white noise, there has a chance that its sub-series may be  white noise. To
examine this, we then apply all tests to five 2-year sub-series, and the results reported in Panel B of Table
\ref{sp500} indicate that both 2012-2013 and 2014-2015 sub-series are white noises at the level 5\%, while the other three two-year sub-series are not.
For these three non-white-noise sub-series, we further check whether their one-year sub-series are white noises.
The results given in Panel C of Table \ref{sp500} show that among six 1-year sub-series, the 2009, 2010 and 2011 sub-series are indeed white noises at the level 5\%.
In all sub-series study, our MFB tests exhibit much more rejection evidence than the GSM tests, and the AQ test fails to do this for the 2010-2011 sub-series and the
2008 and 2011 sub-series.

Overall, our testing results imply that the S\&P500 return series is not white noise during 2006--2008, while it is white noise during 2009--2015. Since the S\&P500 stock market is relatively more volatile in 2006--2008 than 2009--2015, our findings may indicate that the S\&P500 market is more likely to be inefficient when it is more volatile.

\begin{table*}[h!]
	\caption{Testing results for S\&P500 returns.}\label{sp500}
	\setlength{\tabcolsep}{0.7mm}{
		\begin{tabular}{ccccc ccccc cccc}
			\toprule[2pt]
			Time period&$m$&1&2&3&4&5&$m$&1&2&3&4&5&$AQ$\\
			\midrule
			\multicolumn{14}{c}{Panel A: entire 10-year series}\\
			 2006--2015&$MFB_m^g$&\textbf{0.000}&\textbf{0.000}&\textbf{0.000}&\textbf{0.000}&\textbf{0.000}&$GSM_m^g$&\textbf{0.000}&\textbf{0.000}&\textbf{0.000}&\textbf{0.000}&\textbf{0.000}&\textbf{0.006}\\
            &$MFB_{m}^{\vartriangle}$&\textbf{0.000}&\textbf{0.000}&\textbf{0.000}&\textbf{0.000}&\textbf{0.000}&$GSM_m^\vartriangle$&\textbf{0.000}&\textbf{0.000}&\textbf{0.000}&\textbf{0.000}&\textbf{0.000}&\\
            &$MFB_{m}^{e}$&\textbf{0.000}&\textbf{0.000}&\textbf{0.000}&\textbf{0.000}&\textbf{0.000}&$GSM_m^e$&\textbf{0.000}&\textbf{0.000}&\textbf{0.000}&\textbf{0.000}&\textbf{0.000}&\\\\

            \multicolumn{14}{c}{Panel B: 2-year sub-series}\\
			 2006--2007&$MFB_m^g$&\textbf{0.007}&\textbf{0.020}&\textbf{0.011}&\textbf{0.004}&\textbf{0.002}&$GSM_m^g$&\textbf{0.007}&\textbf{0.023}&\textbf{0.022}&\textbf{0.050}&0.088&\textbf{0.015}\\
            &$MFB_{m}^{\vartriangle}$&\textbf{0.004}&\textbf{0.018}&\textbf{0.009}&\textbf{0.001}&\textbf{0.001}&$GSM_m^\vartriangle$&\textbf{0.004}&\textbf{0.019}&\textbf{0.018}&\textbf{0.047}&0.081&\\
            &$MFB_{m}^{e}$&\textbf{0.005}&\textbf{0.019}&\textbf{0.010}&\textbf{0.002}&\textbf{0.001}&$GSM_m^e$&\textbf{0.005}&\textbf{0.019}&\textbf{0.019}&\textbf{0.048}&0.083&\\\\
            2008--2009&$MFB_m^g$&\textbf{0.002}&\textbf{0.000}&\textbf{0.000}&\textbf{0.003}&\textbf{0.000}&$GSM_m^g$&\textbf{0.002}&\textbf{0.003}&\textbf{0.008}&\textbf{0.017}&\textbf{0.035}&\textbf{0.023}\\
            &$MFB_{m}^{\vartriangle}$&\textbf{0.002}&\textbf{0.000}&\textbf{0.000}&\textbf{0.002}&\textbf{0.000}&$GSM_m^\vartriangle$&\textbf{0.002}&\textbf{0.003}&\textbf{0.007}&\textbf{0.015}&\textbf{0.032}&\\
            &$MFB_{m}^{e}$&\textbf{0.002}&\textbf{0.000}&\textbf{0.001}&\textbf{0.002}&\textbf{0.000}&$GSM_m^e$&\textbf{0.002}&\textbf{0.002}&\textbf{0.007}&\textbf{0.017}&\textbf{0.032}&\\\\
            2010--2011&$MFB_m^g$&\textbf{0.039}&\textbf{0.006}&\textbf{0.003}&\textbf{0.042}&0.059&$GSM_m^g$&\textbf{0.039}&0.116&0.092&0.168&0.219&0.141\\
            &$MFB_{m}^{\vartriangle}$&\textbf{0.040}&\textbf{0.005}&\textbf{0.003}&\textbf{0.048}&\textbf{0.048}&$GSM_m^\vartriangle$&\textbf{0.040}&0.104&0.103&0.132&0.176&\\
            &$MFB_{m}^{e}$&\textbf{0.047}&\textbf{0.005}&\textbf{0.004}&\textbf{0.048}&0.055&$GSM_m^e$&\textbf{0.047}&0.114&0.117&0.141&0.184&\\\\
            2012--2013&$MFB_m^g$&0.610&0.783&0.277&0.090&0.094&$GSM_m^g$&0.610&0.870&0.322&0.166&0.244&0.652\\
            &$MFB_{m}^{\vartriangle}$&0.485&0.571&0.169&0.057&0.064&$GSM_m^\vartriangle$&0.485&0.719&0.193&0.097&0.142&\\
            &$MFB_{m}^{e}$&0.505&0.618&0.199&0.066&0.071&$GSM_m^e$&0.505&0.734&0.211&0.118&0.163&\\\\
            2014--2015&$MFB_m^g$&0.406&0.051&0.119&0.229&0.213&$GSM_m^g$&0.406&0.071&0.134&0.322&0.236&0.608\\
            &$MFB_{m}^{\vartriangle}$&0.329&0.076&0.106&0.185&0.170&$GSM_m^\vartriangle$&0.329&0.072&0.112&0.245&0.200&\\
            &$MFB_{m}^{e}$&0.346&0.087&0.109&0.202&0.198&$GSM_m^e$&0.346&0.082&0.128&0.266&0.204&\\\\

            \multicolumn{14}{c}{Panel C: 1-year sub-series}\\
            2006&$MFB_m^g$&0.777&\textbf{0.008}&0.053&0.068&0.065&$GSM_m^g$&0.777&\textbf{0.037}&0.082&0.080&\textbf{0.034}&\textbf{0.002}\\
            &$MFB_{m}^{\vartriangle}$&0.636&\textbf{0.025}&0.064&0.079&0.075&$GSM_m^\vartriangle$&0.636&\textbf{0.048}&0.084&0.090&\textbf{0.050}&\\
            &$MFB_{m}^{e}$&0.644&\textbf{0.032}&0.068&0.079&0.080&$GSM_m^e$&0.644&0.051&0.087&0.080&0.057&\\\\
            2007&$MFB_m^g$&\textbf{0.006}&\textbf{0.041}&0.122&0.050&\textbf{0.037}&$GSM_m^g$&\textbf{0.006}&\textbf{0.018}&\textbf{0.021}&\textbf{0.044}&0.078&\textbf{0.008}\\
            &$MFB_{m}^{\vartriangle}$&\textbf{0.016}&0.057&0.132&0.058&0.052&$GSM_m^\vartriangle$&\textbf{0.016}&\textbf{0.021}&\textbf{0.029}&	 0.058&0.095&\\
            &$MFB_{m}^{e}$&\textbf{0.018}&0.062&0.144&0.059&0.064&$GSM_m^\vartriangle$&\textbf{0.018}&\textbf{0.029}&\textbf{0.035}&0.068&0.110&\\\\
            2008&$MFB_m^g$&\textbf{0.017}&\textbf{0.000}&\textbf{0.000}&\textbf{0.007}&\textbf{0.000}&$GSM_m^g$&\textbf{0.017}&\textbf{0.012}&\textbf{0.031}&0.060&0.105&0.060\\
            &$MFB_{m}^{\vartriangle}$&\textbf{0.025}&\textbf{0.002}&\textbf{0.005}&\textbf{0.011}&\textbf{0.002}&$GSM_m^\vartriangle$&\textbf{0.025}&\textbf{0.020}&\textbf{0.043}&0.068&0.129&\\
            &$MFB_{m}^{e}$&\textbf{0.037}&\textbf{0.002}&\textbf{0.004}&\textbf{0.012}&\textbf{0.002}&$GSM_m^e$&\textbf{0.037}&\textbf{0.024}&0.058&0.078&0.137&\\\\
            2009&$MFB_m^g$&0.089&0.390&0.589&0.707&0.271&$GSM_m^g$&0.089&0.222&0.202&0.279&0.404&0.100\\
            &$MFB_{m}^{\vartriangle}$&0.074&0.368&0.499&0.726&0.283&$GSM_m^\vartriangle$&0.074&0.154&0.127&0.282&0.371&\\
            &$MFB_{m}^{e}$&0.088&0.406&0.535&0.732&0.297&$GSM_m^e$&0.088&0.170&0.145&0.291&0.413&\\\\
            2010&$MFB_m^g$&0.449&0.897&0.916&0.381&0.856&$GSM_m^g$&0.449&0.742&0.857&0.933&0.958&0.413\\
            &$MFB_{m}^{\vartriangle}$&0.373&0.739&0.832&0.302&0.823&$GSM_m^\vartriangle$&0.373&0.507&0.618&0.721&0.753&\\
            &$MFB_{m}^{e}$&0.396&0.745&0.846&0.311&0.845&$GSM_m^e$&0.396&0.514&0.632&0.750&0.758&\\\\
            2011&$MFB_m^g$&0.060&\textbf{0.003}&\textbf{0.000}&\textbf{0.005}&\textbf{0.018}&$GSM_m^g$&0.060&0.163&0.132&0.229&0.315&0.205\\
            &$MFB_{m}^{\vartriangle}$&0.076&\textbf{0.005}&\textbf{0.004}&\textbf{0.016}&\textbf{0.034}&$GSM_m^\vartriangle$&0.076&0.175&0.134&0.235&0.321&\\
            &$MFB_{m}^{e}$&0.080&\textbf{0.007}&\textbf{0.007}&\textbf{0.018}&\textbf{0.042}&$GSM_m^e$&0.080&0.186&0.141&0.238&0.322&\\

			\bottomrule[2pt]
		\end{tabular}
\begin{tablenotes}
      \item[1] {Note: The p-value of each test statistic less than 5\% is in boldface.}
  \end{tablenotes}
	}
\end{table*}

\subsection{Application 2}
This subsection re-visits daily stock returns of BTC,
CCME, KV-A, and MCBF in Francq and Zako\"{i}an (2012).
These four data sets range from  June 29, 2007, March 31, 2009,  March 31, 2006, and August 28, 2007, respectively, to February 7, 2011,
with 907, 468, 1220, and 867, respectively, observations in total.
In Francq and Zako\"{i}an (2012),
all four stock return series are fitted by the non-stationary GARCH($1, 1$) model, while no investigation is given to check whether there exists
serial dependence in their conditional mean. Intuitively, if these four stock return series are white noises,
they can be directly fitted by the non-stationary GARCH($1, 1$) model, otherwise, they possibly have some conditional mean dynamics, which need be filtered out first.

We use our three MFB tests as well as three GSM tests and the automatic portmanteau test $AQ$ to examine whether these four stock return series
are white noises. The testing results are summarized in Table \ref{stocks}, from which we find that only CCME return series is white noise, while the other three return series are not at the level 5\%. Specifically, our MFB tests get more rejection evidence than the GSM tests for the KV-A return series, and the GSM tests do it better especially at the scales $m=3$ and $4$ for the BTC return series. For the
MCBF return series, the white noise hypothesis is strongly rejected by all tests. Compared with the MBF and GSM tests, the test $AQ$ can not find the significant evidence of rejection for BTC, CCME and KV-A return series.

In summary, our testing results imply that
only CCME return series has no serial dependence on its conditional mean, and it is thus suitable to fit this series by the non-stationary GARCH($1, 1$) model. However, the other three return series (particularly, MCBF) most likely have serial dependence on their conditional mean, and without filtering out the conditional mean effect ahead,
the fittings in Francq and Zako\"{i}an (2012) may be inappropriate for these three series.

\begin{table*}[h!]
	\caption{Testing results for four stock returns.}\label{stocks}
	\setlength{\tabcolsep}{0.8mm}{
		\begin{tabular}{ccccc ccccc cccc}
			\toprule[2pt]
			Series&$m$&1&2&3&4&5&$m$&1&2&3&4&5&$AQ$\\
			\midrule
			 BTC&$MFB_m^g$&\textbf{0.001}&\textbf{0.014}&0.053&0.261&\textbf{0.019}&$GSM_m^g$&\textbf{0.001}&\textbf{0.006}&\textbf{0.012}&\textbf{0.019}&\textbf{0.027}&0.052\\
			 &$MFB_{m}^{\vartriangle}$&\textbf{0.001}&\textbf{0.011}&\textbf{0.042}&0.113&\textbf{0.012}&$GSM_m^\vartriangle$&\textbf{0.001}&\textbf{0.004}&\textbf{0.011}&\textbf{0.012}&\textbf{0.020}\\
			 &$MFB_{m}^{e}$&\textbf{0.002}&\textbf{0.014}&\textbf{0.047}&0.133&0.016&$GSM_m^e$&\textbf{0.002}&\textbf{0.004}&\textbf{0.012}&\textbf{0.016}&\textbf{0.022}\\\\
			
			CCME&$MFB_m^g$&0.622&0.659&0.279&0.162&0.447&$GSM_m^g$&0.622&0.699&0.296&0.437&0.545&0.814\\
			&$MFB_{m}^{\vartriangle}$&0.543&0.557&0.122&0.080&0.361&$GSM_m^\vartriangle$&0.543&0.589&0.152&0.354&0.436\\
			&$MFB_{m}^{e}$&0.576&0.580&0.149&0.089&0.379&$GSM_m^e$&0.576&0.604&0.177&0.370&0.454\\\\
			
			KV-A&$MFB_m^g$&0.111&0.088&0.077&0.110&0.086&$GSM_m^g$&0.111&0.042&0.094&0.170&0.267&0.347\\
			 &$MFB_{m}^{\vartriangle}$&0.061&\textbf{0.044}&\textbf{0.043}&0.060&\textbf{0.043}&$GSM_m^\vartriangle$&0.061&\textbf{0.046}&\textbf{0.050}&0.091&0.117\\
			&$MFB_{m}^{e}$&0.061&\textbf{0.048}&\textbf{0.049}&0.059&\textbf{0.047}&$GSM_m^e$&0.061&\textbf{0.047}&0.052&0.096&0.125\\\\
			
			 MCBF&$MFB_m^g$&\textbf{0.000}&\textbf{0.000}&\textbf{0.000}&\textbf{0.000}&\textbf{0.000}&$GSM_m^g$&\textbf{0.000}&\textbf{0.000}&\textbf{0.000}&\textbf{0.000}&\textbf{0.000}&\textbf{0.001}\\
			 &$MFB_{m}^{\vartriangle}$&\textbf{0.000}&\textbf{0.000}&\textbf{0.000}&\textbf{0.000}&\textbf{0.000}&$GSM_m^\vartriangle$&\textbf{0.000}&\textbf{0.000}&\textbf{0.000}&\textbf{0.000}&\textbf{0.000}\\
			 &$MFB_{m}^{e}$&\textbf{0.000}&\textbf{0.000}&\textbf{0.000}&\textbf{0.000}&\textbf{0.000}&$GSM_m^e$&\textbf{0.000}&\textbf{0.000}&\textbf{0.000}&\textbf{0.000}&\textbf{0.000}\\

			\bottomrule[2pt]
		\end{tabular}
\begin{tablenotes}
      \item[1] {Note: The p-value of each test statistic less than 5\% is in boldface.}
  \end{tablenotes}
	}
\end{table*}



\appendix

\section*{Appendix: Technical conditions and proofs}

To introduce our technical conditions, the definition of \textit{near-epoch dependence} is needed.

\begin{de}
	For a stochastic sequence $\{\epsilon_t\}$, let $\mathcal{F}^{t+m}_{t-m}(\epsilon)=\sigma(\epsilon_{t-m},...,\epsilon_{t+m})$. A stochastic sequence $\{y_t\}$ is near-epoch dependent (NED) on $\{\epsilon_t\}$ in $L_p$-norm for $p>0$ if
	\[{\left\| y_t-E[y_t|\mathcal{F}^{t+m}_{t-m}(\epsilon)]\right\|}_p\leq d_t\nu_m, \]
	where $\nu_m\to0$ as $m\to\infty$, and $d_t$ is a sequence of positive real numbers such that $d_t=O(\left\|  y_t\right\|_p )$.
\end{de}

The concept of near-epoch dependence can be traced back to the work of Ibragimov (1962). The NED processes allow for considerable
heterogeneity and also for dependence and include the mixing processes as
a special case. As shown in Davidson
(2002, 2004) and references therein, many nonlinear models are shown to be NED.

Next, we are ready to give our technical conditions.

\begin{ass}\label{ass1}
$\{y_t\}$ is a stochastic process which is $L_r$-bounded for $r>2$ and $L_p$-NED on an $\alpha$-mixing process for $p\geq2$.
\end{ass}

\begin{ass}\label{ass2}
	(i) For $r>1$ and for all $i,j,k,l$ such that $0\leq i<j\leq L_m$ and $0\leq k<l\leq L_m$, $\{y_{t-i}y_{t-j}y_{t-k}y_{t-l}/M^y_{4,t}\}$ is uniformly $L_r$-bounded for $r>1$, where
\[M^y_{4,t}=\sum_{i=0}^{L_m}\sum_{j>1}^{L_m}\sum_{k=0}^{L_m}\sum_{l>1}^{L_m}\widetilde{v}_{m,n,i}\widetilde{v}_{m,n,j}\widetilde{v}_{m,n,k}\widetilde{v}_{m,n,l} E(y_{t-i}y_{t-j}y_{t-k}y_{t-l}).\]
	
 (ii) For all positive $i\leq L_m$, $\{y_{t}y_{t-j}\}$ is a $L_r$-bounded stochastic sequence for $r>2$ and $L_p$-NED of size $-1/2$ on a $\phi$-mixing process $\{\epsilon_t\}$ for $p\geq2$.
		
 (iii) $\mbox{var}(z_{m,n,t})\sim t^\beta$ and $s_{m,n,T}^2(z)\sim T^{1+\gamma}$ for $\beta\leq\gamma$.
\end{ass}

Assumptions \ref{ass1}--\ref{ass2} are in line with Assumptions A--B in Gen\c{c}ay and Signori (2015), and they
allow for the heteroskedastic data.
For the GARCH(1, 1) model, the NED conditions in Assumptions \ref{ass1}--\ref{ass2} were verified by
Gen\c{c}ay and Signori (2015). For the general model, it seems challenging to verify Assumptions \ref{ass1}--\ref{ass2} in theory at this stage. Nevertheless,
the good finite-sample performance of our MFB tests in Section 4 implies that
these two assumptions could hold for a variety of time series models.

\subsection*{Proof of Theorem \ref{t1}}
According to the construction of MODWPT, $W_{m,n,t}$ can be obtained by applying the filter $\{\widetilde{v}_{m,n,l}\}$ to the process $\{y_t\}$, where $\{\widetilde{v}_{m,n,l}\}$ only depends on $\{h_l\}$ and $\{g_l\}$. Let $V_{m,n}(\cdot)$ be the discrete Fourier transfer function for $\{\widetilde{v}_{m,n,l}\}$, which depends only on the transfer functions $G_m(\cdot)$ and $H_m(\cdot)$ for $\{h_l\}$ and $\{g_l\}$, respectively (see, e.g., the specific expressions in Percival and Walden (2000, p.215)). Then,  when $\{y_t\}$ is stationary, the spectrum of $W_{m,n,t}$ is $S_{W_{m,n}}(\cdot)=\left|V_{m,n}(\cdot)\right|^2S_y(\cdot)$, and
since $S_y(f)=\sigma^2_y$ for a covariance stationary white noise $\{y_t\}$, it follows that
\begin{align}\label{2e3}
{\rm var}(W_{m,n,t})&=\int^{\frac{1}{2}}_{-\frac{1}{2}}S_{W_{m,n}}(f)\mathrm{d}f
=\int^{\frac{1}{2}}_{-\frac{1}{2}}\left|V_{m,n}(f)\right|^2S_y(f)\mathrm{d}f
=\sigma^2_y\int^{\frac{1}{2}}_{-\frac{1}{2}}\left|V_{m,n}(f)\right|^2\mathrm{d}f\nonumber\\
&=\sigma^2_y\left\|\widetilde{v}_{m,n}\right\|_2
=\sigma^2_y\left\|g\right\|_2^{i_{m,n}}\left\|h\right\|_2^{m-i_{m,n}}
=\sigma^2_y/2^m,
\end{align}
where (\ref{2e3}) holds by Parseval's identity and the basic properties of the wavelet filter and its associated scaling filters, and $i_{m,n}$ is an integer satisfying $0\leq i_{m,n}\leq m$, which is only determined by $m$ and $n$ (see Percival and Walden (2000, p.215)). $\Box$

\subsection*{Proof of Theorem \ref{t2}}
(i) Since the NED property is preserved under linear combinations (see Davidson (1995, p.267)) and the MODWPT is a linear operator, $\{W_{m,n,t}\}$ is $L_2$-NED under Assumption \ref{ass1}, and consequently, $\{W^2_{m,n,t}\}$ is $L_1$-NED (see Davidson (1995, p.268)), where
\begin{equation*}
W^2_{m,n,t}=\sum_{l=0}^{L_m-1}\widetilde{v}^2_{m,n,l}y^2_{t-l}
+2\sum_{i=0}^{L_m-1}\sum_{j>i}^{L_m}\widetilde{v}_{m,n,i}\widetilde{v}_{m,n,j}y_{t-i}y_{t-j}=\sum_{l=1}^{L_m}\widetilde{v}^2_{m,n,l}y^2_{t-l}+2z_{m,n,t},
\end{equation*}
and $\{z_{m,n,t}\}$ is $L_1$-NED because it is a linear combination of $\{W^2_{m,n,t}\}$ and $\{y^2_t\}$, both of which are $L_1$-NED. Then, it follows that
\begin{align}
\widehat{\xi}_{m,n,T}&=\f{\sum^{T}_{t=1}W^2_{m,n,t}}{\sum^{T}_{t=1}y^2_t}=\f{\sum^{T}_{t=1}\left( \sum_{l=0}^{L_m-1}\widetilde{v}^2_{m,n,l}y^2_{t-l}+2z_{m,n,t}\right) }{\sum^{T}_{t=1}y^2_t}\nonumber\\
&=\f{ \sum_{l=0}^{L_m-1}\widetilde{v}^2_{m,n,l}\sum^{T}_{t=1}y^2_{t-l}}{\sum^{T}_{t=1}y^2_t}+\f{2\sum_{t=1}^{T}z_{m,n,t}}{\sum^{T}_{t=1}y^2_t}\nonumber\\
&=\sum_{l=0}^{L_m-1}\widetilde{v}^2_{m,n,l}+\f{2\sum_{t=1}^{T}z_{m,n,t}}{\sum^{T}_{t=1}y^2_t} \label{A2}\\
&=\f{1}{2^m}+\f{2\sum_{t=1}^{T}z_{m,n,t}}{\sum^{T}_{t=1}y^2_t},\label{A3} 
\end{align}
where (\ref{A2}) holds since the filtering is cyclic so that $\sum^{T}_{t=1}y^2_{t-l}$ is not related to $l$ and is equal to $\sum^{T}_{t=1}y^2_{t}$,
and (\ref{A3}) holds since for $m$th level of MODWPT, each of $\{\widetilde{v}_{m,n,t}\}_{n=0}^{2^m-1}$ is the cascade filters obtained by convolution of $m$ filters with norm $1/2$, and the norm of a convolution is the product of the norms. Finally, the conclusion holds since
\begin{equation*}
\f{2\sum_{t=1}^{T}z_{m,n,t}}{\sum^{T}_{t=1}y^2_t}\xrightarrow{p}0\mbox{ as }T\to\infty
\end{equation*}
by Theorem 1 of Andrews (1988) and Slutsky's Theorem.

(ii)
Since the NED property is preserved under linear combinations and $\{z_{m,n,t}\}$ is a linear combination of processes of the form $\{y_ty_{t-i}\}$, we can get that $\{z_{m,n,t}\}$ is $L_r$-NED on $\{\epsilon_t\}$ under Assumption \ref{ass2}.
Next, we will verify that $\{z_{m,n,t}\}$ satisfies the conditions of the Central Limit Theorem for NED processes in De Jong (1997, p.358). Note that
\begin{align*}
{\rm var}(z_{m,n,t})&={\rm var}\left( \sum_{i=0}^{L_m-1}\sum_{j>i}^{L_m}\widetilde{v}_{m,n,i}\widetilde{v}_{m,n,j}y_{t-i}y_{t-j}\right) \\
&={\rm cov}\left(\sum_{i=0}^{L_m-1}\sum_{j>i}^{L_m}\widetilde{v}_{m,n,i}\widetilde{v}_{m,n,j}y_{t-i}y_{t-j},
\sum_{k=0}^{L_m-1}\sum_{l>k}^{L_m}\widetilde{v}_{m,n,k}\widetilde{v}_{m,n,l}y_{t-k}y_{t-l}\right)\\
&=\sum_{i=0}^{L_m-1}\sum_{j>i}^{L_m}\sum_{k=0}^{L_m-1}\sum_{l>k}^{L_m}\widetilde{v}_{m,n,i}\widetilde{v}_{m,n,j}\widetilde{v}_{m,n,k}
\widetilde{v}_{m,n,l}{\rm cov}\left(y_{t-i}y_{t-j},y_{t-k}y_{t-l}\right)\\
&=\sum_{i=0}^{L_m-1}\sum_{j>i}^{L_m}\sum_{k=0}^{L_m-1}\sum_{l>k}^{L_m}\widetilde{v}_{m,n,i}\widetilde{v}_{m,n,j}\widetilde{v}_{m,n,k}
\widetilde{v}_{m,n,l}{\rm E}\left(y_{t-i}y_{t-j}y_{t-k}y_{t-l}\right),
\end{align*}
where the last equation holds because the mean of $\{y_t\}$ is zero. Then, we have
\begin{align*}
&\left\|\f{y_{t-i}y_{t-j}y_{t-k}y_{t-l}}{\sum_{i=0}^{L_m-1}\sum_{j>i}^{L_m}\sum_{k=0}^{L_m-1}\sum_{l>k}^{L_m}
\widetilde{v}_{m,n,i}\widetilde{v}_{m,n,j}\widetilde{v}_{m,n,k}\widetilde{v}_{m,n,l}{\rm E}\left(y_{t-i}y_{t-j}y_{t-k}y_{t-l}\right)}\right\|_p\\
&\sim\left\|\f{\sum_{i=0}^{L_m-1}\sum_{j>i}^{L_m}\sum_{k=0}^{L_m-1}\sum_{l>k}^{L_m}
\widetilde{v}_{m,n,i}\widetilde{v}_{m,n,j}\widetilde{v}_{m,n,k}\widetilde{v}_{m,n,l}y_{t-i}y_{t-j}y_{t-k}y_{t-l}}{\sum_{i=0}^{L_m-1}\sum_{j>i}^{L_m}
\sum_{k=0}^{L_m-1}\sum_{l>k}^{L_m}\widetilde{v}_{m,n,i}\widetilde{v}_{m,n,j}\widetilde{v}_{m,n,k}\widetilde{v}_{m,n,l}{\rm E}\left(y_{t-i}y_{t-j}y_{t-k}y_{t-l}\right)}\right\|_p\\
&=\left\|\f{z_{m,n,t}^2}{{\rm var}(z_{m,n,t})}\right\|_p
=\left\|\f{z_{m,n,t}}{\sigma_{m,n,t}}\right\|_{2p},
\end{align*}
which implies that $z_{m,n,t}/\sigma_{m,n,t}$ is $L_{q}$-bounded for $q=2p>2$. Hence, we have verified that $\{z_{m,n,t}\}$ satisfies the conditions of the Central Limit Theorem for NED processes, and so we have
\begin{equation*}
\sum_{t=1}^T\f{z_{m,n,t}}{s_{m,n,T}(z)}\xrightarrow{d}N(0,1)\mbox{ as }T\to\infty.
\end{equation*}
By (\ref{A3}), it follows that
\begin{equation*}
\f{\sum_{t=1}^Ty_t^2}{2s_{m,n,T}(z)}\left(\widehat{\xi}_{m,n,T}-\f{1}{2^m}\right)\xrightarrow{d}N(0,1)\mbox{ as }T\to\infty.
\end{equation*}
Since $\frac{1}{T}\sum_{t=1}^{T}Ey_t^2\rightarrow\sigma^2$ as $T\to\infty$, the conclusion follows by Slutsky's Theorem.
$\Box$

\subsection*{Proof of Proposition \ref{p1}}
The conclusion holds by the similar arguments as for Corollary 13 in Gen\c{c}ay and Signori (2015), and hence the details are omitted. $\Box$

\end{document}